\documentclass[aps,prd,amsmath,floatfix,twocolumn,superscriptaddress,nofootinbib,showpacs]{revtex4-1}
\usepackage{graphicx}
\usepackage{amsfonts}
\usepackage{subfigure}
\usepackage{xspace}
\usepackage[usenames,dvipsnames]{color}
\usepackage{dcolumn}
\usepackage{bm}
\usepackage{mathrsfs}
\usepackage[]{amsmath,amssymb}
\usepackage[]{amsthm}
\usepackage{verbatim}
\usepackage{appendix}
\usepackage[colorinlistoftodos]{todonotes}
\usepackage{mathrsfs}
\usepackage{mathtools}
\usepackage{comment}
\usepackage{tikz}
\usepackage{tikz-3dplot}
\usepackage{pgfplots}
\usepackage{pgf}

\definecolor{CiteColor}{rgb}{0.18039, 0.18824, 0.57255}
\definecolor{UrlColor} {rgb}{0.741, 0.173, 0.000}
\definecolor{LinkColor}{rgb}{0.25098, 0.47843, 0.04706}
\usepackage[colorlinks, plainpages=false,
hyperfigures=true]{hyperref}
\hypersetup{linkcolor=LinkColor}
\hypersetup{citecolor=CiteColor}
\hypersetup{urlcolor=UrlColor}

\usepackage{ulem}
\newcommand{\Caltech}{\affiliation{Theoretical Astrophysics 350-17,
    California Institute of Technology, Pasadena, CA 91125}} %
\newcommand{\Cornell}{\affiliation{Center for Radiophysics and Space
    Research, Cornell University, Ithaca, New York, 14853}} %
\newcommand{\CIFAR}{\affiliation{Canadian Institute for Advanced
    Research, 180 Dundas St.~West, Toronto, ON M5G 1Z8, Canada}} %
\newcommand{\DAA}{\affiliation{Department of Astronomy and
    Astrophysics, 50 St.\ George Street, University of Toronto,
    Toronto, ON M5S 3H4, Canada}}

\newcommand{\CITA}{\affiliation{Canadian Institute for Theoretical
    Astrophysics, University~of~Toronto, Toronto, Ontario M5S 3H8,
    Canada}} %
\newcommand{\abs}[1]{\lvert#1\rvert}

\newcommand{\ellHat}{\ensuremath{\hat{\ell}}}
\newcommand{\nHat}{\ensuremath{\hat{n}}}
\newcommand{\lambdaHat}{\ensuremath{\hat{\lambda}}}

\newcommand{\Rf}[1][]{\ensuremath{R_{\text{f#1}}}}
\newcommand{\Rbarf}[1][]{\ensuremath{\bar{R}_{\text{f#1}}}}

\newcommand{\RfA}{\Rf[\hspace{0.1em}A]}
\newcommand{\RfB}{\Rf[\hspace{0.1em}B]}
\newcommand{\RbarfA}{\Rbarf[\hspace{0.1em}A]}

\newcommand{\RfPN}{\Rf^{\text{PN}}}
\newcommand{\RfNR}{\Rf^{\text{NR}}}

\newcommand{\RbarfNR}{\Rbarf^{\text{NR}}}

\newcommand{\U}{\ensuremath{\mathrm{U}}}
\newcommand{\SU}{\ensuremath{\mathrm{SU}}}
\newcommand{\SO}{\ensuremath{\mathrm{SO}}}

\newcommand{\defined}{\coloneqq}

\allowdisplaybreaks

\begin{document}

\title{Comparing Post-Newtonian and Numerical-Relativity Precession
  Dynamics}

\author{{Serguei Ossokine}}\CITA \DAA
\author{{Michael Boyle}} \Cornell 
\author{{Lawrence~E.~Kidder}}\Cornell 
\author{{Harald P. Pfeiffer}}\CITA \CIFAR
\author{{Mark A.~Scheel}} \Caltech 
\author{B\'{e}la Szil\'{a}gyi} \Caltech 

\begin{abstract}
  Binary black-hole systems are expected to be important sources of
  gravitational waves for upcoming gravitational-wave detectors.  If
  the spins are not colinear with each other or with the orbital
  angular momentum, these systems exhibit complicated precession
  dynamics that are imprinted on the gravitational waveform.  We
  develop a new procedure to match the precession dynamics computed by
  post-Newtonian (PN) theory to those of numerical binary black-hole
  simulations in full general relativity.  For numerical relativity
  (NR) simulations lasting approximately two precession cycles, we
  find that the PN and NR predictions for the directions of the
  orbital angular momentum and the spins agree to better than $\sim
  1^{\circ}$ with NR during the inspiral, increasing to $5^{\circ}$
  near merger.  Nutation of the orbital plane on the orbital
  time-scale agrees well between NR and PN, whereas nutation of the
  spin direction shows qualitatively different behavior in PN and NR.
  We also examine how the PN equations for precession and
  orbital-phase evolution converge with PN order, and we quantify the
  impact of various choices for handling partially known PN terms.
\end{abstract}
\date{\today}
\maketitle


\section{Introduction}
\label{sec:Introduction}

\begin{figure*}
\includegraphics[width=0.98\linewidth,trim=12 24 12 35,clip=true]{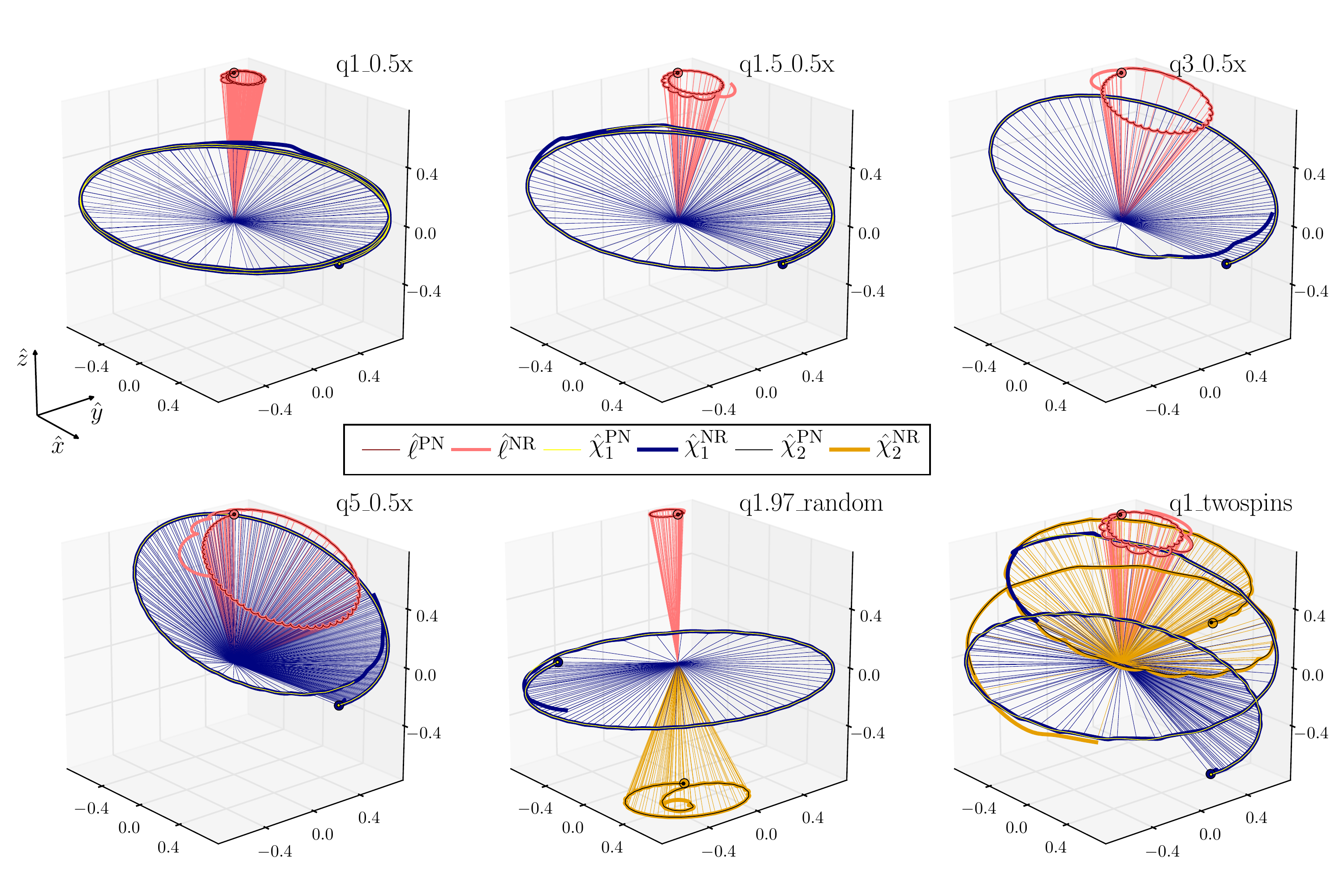}
\caption{Precession cones of the six primary precessing simulations
  considered here, as computed by NR and PN.  Shown are the paths
  traced on the unit sphere by the normal to the orbital plane
  $\ellHat$ and the spin-directions $\hat{\chi}_{1,2}$. The thick
  lines represent the NR data, with the filled circles indicating the
  start of the NR simulations.  The lines connecting the NR data to
  the origin are drawn to help visualize the precession-cones.  The PN
  data, plotted with thin lines, lie on the scale of this figure
  almost precisely on top of the NR data.  (The PN data was
  constructed using the Taylor T4 approximant matched at frequency
  $m\Omega_m=0.021067$, with a matching interval width
  $\delta\Omega=0.1\Omega_m$.) }
  \label{fig:precessionCones}
\end{figure*}

Binary black holes (BBH) are among the most important sources of
gravitational waves for upcoming gravitational-wave detectors like
Advanced LIGO~\cite{TheLIGOScientific:2014jea} and
Virgo~\cite{Accadia:2011zzc}.  Accurate predictions of the
gravitational waveforms emitted by such systems are important for
detection of gravitational waves and for parameter estimation of any
detected binary~\cite{Abbott:2007}.  When either black hole carries
spin that is \emph{not} aligned with the orbital angular momentum,
there is an exchange of angular momentum between the components of the
system, leading to complicated dynamical behavior.
Figure~\ref{fig:precessionCones} exhibits the directions of the
various angular momenta in several simulations described in this
paper.  This behavior is imprinted on the emitted
waveforms~\cite{Apostolatos1994, PekowskyEtAl:2013, BoyleEtAl:2014},
making them more feature-rich than waveforms from aligned-spin BBH
systems or non-spinning BBH systems.  In order to model the waveforms
accurately, then, we need to understand the dynamics.

The orbital-phase evolution of an inspiraling binary, the precession
of the orbital angular momentum and the black-hole spins, and the
emitted gravitational waveforms can be modeled with post-Newtonian
theory~\cite{lrr-2014-2}, a perturbative solution of Einstein's
equations in powers of $v/c$, the ratio of the velocity of the black
holes to the speed of light.  Such post-Newtonian waveforms play an
important role in the waveform modeling for ground-based
interferometric gravitational-wave detectors (see,
e.g.,~\cite{Ohme:2011rm}).  For non-spinning and aligned-spin BBH,
however, the loss of accuracy of the post-Newtonian phase evolution in
the late inspiral has been identified as one of the dominant
limitations of waveform modeling~\cite{Damour:2010, Boyle:2011dy,
  OhmeEtAl:2011, MacDonald:2011ne, MacDonald:2012mp, Nitz:2013mxa}.

Precessing waveform models (e.g.,~\cite{Hannam:2013oca,
  Taracchini:2013rva, Pan:2013rra, BoyleEtAl:2014}) depend on the
orbital phase evolution and the precession dynamics.  Therefore, it is
important to quantify the accuracy of the post-Newtonian approximation
for modeling the precession dynamics itself, and the orbital-phase
evolution of precessing binaries.  Recently, the SXS collaboration has
published numerical-relativity solutions to the full Einstein
equations for precessing BBH systems~\cite{PhysRevLett.111.241104}.
These simulations cover $\gtrsim 30$ orbits and up to two precession
cycles.  Therefore, they offer a novel opportunity to systematically
quantify the accuracy of the post-Newtonian precession equations, the
topic of this paper.

In this paper, we develop a new technique to match the initial
conditions of post-Newtonian dynamics to a numerical relativity
simulation.  We then use this technique to study the level of
agreement between the post-Newtonian precession equations and the
numerical simulations.  The agreement is remarkably good, the
directions of orbital angular momentum and spin axes in post-Newtonian
theory reproduces the numerical simulations usually to better than $1$
degree.  We also investigate nutation effects on the orbital
time-scale that are imprinted both in the orbital angular momentum and
the spin-directions.  For the orbital angular momentum, NR and PN
yield very similar nutation features, whereas for the spin direction,
nutation is qualitatively different in PN and the investigated NR
simulations.  Considering the orbital-phase evolution, we find that
the disagreement between post-Newtonian orbital phase and numerical
relativity simulation is comparable to the aligned-spin case. This
implies that the orbital phase evolution will remain an important
limitation for post-Newtonian waveforms even in the precessing case.
Finally, we study the convergence with post-Newtonian order of the
precession equations, and establish very regular and fast convergence,
in contrast to post-Newtonian orbital phasing.

This paper is organized as follows: Section~\ref{sec:Methodology}
describes the post-Newtonian expressions utilized, the numerical
simulations, how we compare PN and NR systems with each other, and how
we determine suitable ``best-fitting'' PN parameters for a comparison
with a given NR simulation.  Section~\ref{sec:comparison} presents our
results, starting with a comparison of the precession dynamics
in Sec.~\ref{sec:PrecessionComparisons}, and continuing with an
investigation in the accuracy of the orbital phasing in
Sec.~\ref{sec:OrbitalPhaseComparisons}.  The following two sections
study the convergence of the PN precession equations and the impact of
ambiguous choices when dealing with incompletely known spin-terms in
the PN orbital phasing.  Section~\ref{sec:tech_cons}, finally, is
devoted to some technical numerical aspects, including an
investigation into the importance of the gauge conditions used for the
NR runs.  We close with a discussion in Sec.~\ref{sec:discussion}.
The appendices collect the precise post-Newtonian expressions we use
and additional useful formulae about quaternions.

\section{Methodology}
\label{sec:Methodology}


\subsection{Post-Newtonian Theory}
\label{sec:PostNewtonianTheory}

Post-Newtonian (PN) theory is an approximation to General Relativity
in the weak-field, slow-motion regime, characterized by the small
parameter $\epsilon \sim (v/c)^{2} \sim \frac{Gm}{rc^2}$, where $m$,
$v$, and $r$ denote the characteristic mass, velocity, and size of the
source, $c$ is the speed of light, and $G$ is Newton's gravitational
constant.  For the rest of this paper, the source is always a binary
black-hole system with total mass $m$, relative velocity $v$ and
separation $r$, and we use units where $G=c=1$.

Restricting attention to quasi-spherical binaries in the
adiabatic limit, the local dynamics of the source can be split into
two parts: the evolution of the orbital frequency, and the precession
of the orbital plane and the spins. The leading-order precessional
effects~\cite{Barker:1975ae} and spin contributions to the evolution
of the orbital frequency~\cite{kidder93,Kidder:1995zr} enter
post-Newtonian dynamics at the 1.5~PN order (i.e., $\epsilon^{3/2}$)
for spin-orbit effects, and 2~PN order for spin-spin effects. We also
include non-spin terms to 3.5~PN order \cite{lrr-2014-2}, the
spin-orbit terms to 4~PN order \cite{Marsat:2013caa}, spin-spin terms
to 2~PN order \cite{Kidder:1995zr}\footnote{ During the
    preparation of this manuscript, the 3 PN spin-spin contributions
    to the flux and binding energy were completed in
    \cite{Bohe:2015ana}. These terms are not used in the analysis
    presented here.}. For the precession equations, we include the
spin-orbit contributions to next-to-next-to-leading order,
corresponding to 3.5~PN \cite{BoheEtAl:2013}. The spin-spin terms are
included at 2~PN order\footnote{The investigation of the
    effects of spin-spin terms at higher PN~orders (see
    e.g. \cite{Hartung:2011ea,PR08a,Levi:2014sba} and references
    therein), and terms which are higher order in spin (e.g cubic spin
    terms) \cite{Marsat:2014xea,Levi:2014gsa} is left for future
    work.}.


\subsubsection{Orbital dynamics}
\label{sec:OrbitalEvolution}

\begin{figure}
  \includegraphics[width=0.96\linewidth]{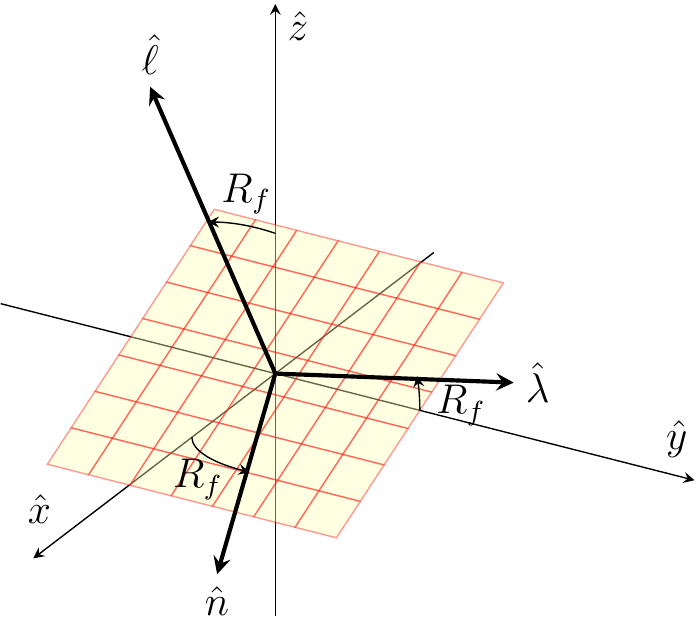}
  \caption{Vectors describing the orbital dynamics of the system. The
    yellow plane denotes the orbital plane.  $\Rf(t)$ is the rotor
    that rotates the coordinate triad $(\hat{x}, \hat{y}, \hat{z})$
    into the orbital triad $(\nHat,\lambdaHat,\ellHat)$.}
  \label{fig:orbitalDefns}
\end{figure}

Following earlier work (e.g., Ref.~\cite{Kidder:1995zr}) we describe the
precessing BH binary by the evolution of the orthonormal triad
$(\hat{n},\hat{\lambda},\ellHat)$, as indicated in
Fig.~\ref{fig:orbitalDefns}: $\hat{n}$ denotes the unit separation
vector between the two compact objects, $\ellHat$ is the normal to the
orbital plane and $\lambdaHat=\ellHat\times \nHat$ completes the
triad. This triad is time-dependent, and
is related to the constant inertial triad
$(\hat{x},\hat{y},\hat{z})$ by a time-dependent rotation $R_{f}$, as
indicated in Fig.~\ref{fig:orbitalDefns}. The rotation $R_f$ will play
an important role in Sec.~\ref{CharacterizingPrecessionByRotors}.  The
orbital triad obeys the following
equations:

\begin{subequations}
  \label{eq:TriadEvolution}
  \begin{align}
    \frac{d\ellHat}{dt} &= \varpi\nHat\times\ellHat \label{eq:ellHatEv},\\
    \frac{d\hat{n}}{dt} &= \Omega\hat{\lambda},\label{eq:n-evolution}\\
    \frac{d\hat{\lambda}}{dt} &= -\Omega\hat{n} + \varpi\ellHat. \label{eq:lambEv}
  \end{align}
\end{subequations}
Here, $\Omega$ is the instantaneous orbital frequency and
$\varpi$ is the precession frequency of the orbital
plane.

The dimensionless spin vectors $\vec{\chi}_i = \vec{S}_i/m^2_i$ also
obey precession equations:
\begin{subequations}
  \label{eq:SpinsEv}
  \begin{align}
    \frac{d\vec{\chi}_{1}}{dt} &= \vec{\Omega}_{1}\times \vec{\chi}_{1},
    \\
    \frac{d\vec{\chi}_{2}}{dt} &= \vec{\Omega}_{2}\times \vec{\chi}_{2}.
  \end{align}
\end{subequations}
The precession frequencies $\vec{\Omega}_{1,2},\ \varpi$ are series in
the PN expansion parameter $\epsilon$; their explicit form is given in
Appendix~\ref{ap:PN}.

The evolution of the orbital frequency is derived from energy balance:
\begin{equation}
  \label{eq:eBalance}
  \frac{dE}{dt} = -\mathcal{F},
\end{equation}
where $E$ is the energy of the binary and $\mathcal{F}$ is the
gravitational-wave flux. $E$ and $\mathcal{F}$ are PN series depending
on the orbital frequency $\Omega$, the vector $\ellHat$, and the BH
spins $\vec{\chi}_{1},\ \vec{\chi}_{2}$. Their explicit formulas are
given in Appendix~\ref{ap:PN}. In terms of
  $x\equiv(m\Omega)^{2/3} \sim \epsilon$, Eq.~\eqref{eq:eBalance}
becomes:
\begin{equation}
  \label{eq:OmEv}
  \frac{dx}{dt} = -\frac{\mathcal{F}}{dE/dx},
\end{equation}
where the right-hand side is a ratio of two PN series.

There are several well known ways of solving Eq.~\eqref{eq:OmEv},
which lead to different treatment of uncontrolled higher-order PN
terms---referred to as the Taylor T1 through T5
approximants~\cite{Damour:2000zb,Ajith:2011ec}.  The T2 and T3
approximants cannot be applied to general precessing systems; we
therefore exclude them from this work.  We now briefly review the
remaining approximants, which will be used throughout this
work.\footnote{See, e.g., Ref.~\cite{Boyle2007} for a more complete
  description of approximants T1 through T4.}  The most
straightforward approach is to evaluate the numerator and denominator
of Eq.~\eqref{eq:OmEv} and then solve the resulting ordinary
differential equation numerically, which is the Taylor T1
approximant. Another approach is to re-expand the ratio
$\mathcal{F}/(dE/dx)$ in a new power series in $x$, and then truncate at the
appropriate order. This gives the Taylor T4 approximant. Finally, one
can expand the \emph{inverse of the right-hand-side of
  Eq.~\eqref{eq:OmEv}} in a new power series in $x$, truncate it at the
appropriate order, and then substitute the inverse of the truncated
series into the right-hand side in Eq.~\eqref{eq:OmEv}.  This last
approach, known as the Taylor T5 approximant~\cite{Ajith:2011ec}, has
been introduced fairly recently.

\begin{table*}
  \caption{  \label{tbl:Parameters}Numerical
    relativity simulations utilized here.  SXS ID refers to the
    simulation number in Ref.~\cite{PhysRevLett.111.241104}, $q=m_1/m_2$ is the mass ratio,
    $\vec{\chi}_{1,2}$ are the dimensionless spins, given in
    coordinates where $\nHat(t=0) = \hat{x}$,
    $\ellHat(t=0)=\hat{z}$. $D_{0}$, $\Omega_{0}$ and $e$ are the initial
    coordinate separation, the initial orbital
    frequency, and the orbital eccentricity, respectively.
    The first block  lists the precessing runs utilized, where
    $\vec{\chi}_{1,r}=(-0.18,-0.0479,-0.0378)$ and $\vec{\chi}_{2,r}=
    (-0.0675,0.0779,-0.357)$. 
    The second block indicates 31 further precessing simulations used
    in Fig.~\ref{fig:AngleLRandom31}, and the last block lists the
    aligned spin systems for  orbital phase comparisons.  
  }
  \begin{ruledtabular}
  \begin{tabular}{@{}llllllllll@{}}
     Name &  SXS ID  & $q$ & $\vec{\chi}_{1}$ &
    $\vec{\chi}_{2}$ & $D_{0}/M$ & $m\Omega_{0}$ & $e$ \\
    \hline
    q1\_0.5x &  0003   & 1.0 & (0.5,0.0,0) & (0,0,0) & 19 &
    0.01128 &  0.003 \\
    q1.5\_0.5x & 0017  & 1.5 & (0.5,0,0) & (0,0,0) & 16 &
    0.01443 & $<2\times10^{-4}$ \\
    q3\_0.5x & 0034  & 3.0 & (0.5,0,0) & (0,0,0)  & 14 &
    0.01743 & $<2\times10^{-4}$ \\
    q5\_0.5x &    & 5.0 & (0.5,0,0) & (0,0,0) & 15 &
    0.01579 &  0.002  \\
    q1\_two\_spins & 0163   & 1.0 & (0.52,0,-0.3) & (0.52,0,0.3)
    & 15.3 & 0.01510 & 0.003 \\
    q1.97\_random  & 0146    & 1.97
      & $\vec{\chi}_{1,r}$
      & $\vec{\chi}_{2,r}$
      & 15 & 0.01585 & $<10^{-4}$  \\ 
\\[-0.95em]
\hline
\\[-.95em]
31 random runs & 115--145 & $[1,2]$ & $\chi_1\le 0.5$ & $\chi_2\le 0.5$ & 
15 & $\approx 0.0159$ &  $[10^{-4}, 10^{-3}]$ \\ 
\\[-0.95em]
\hline\\[-.95em]
    \verb|q1_0.5z| & 0005  & 1.0 & (0,0,0.5) & (0,0,0) & 19 &
    0.01217 & 0.0003\\
    \verb|q1_-0.5z| & 0004 & 1.0 & (0,0,0.5) & (0,0,0)  & 19 &
    0.01131 & 0.0004\\
    \verb|q1.5_0.5z| &  0013  & 1.5 & (0,0,0.5) &
    (0,0,0)  & 16 & 0.01438&  0.00014\\
    \verb|q1.5_-0.5z| & 0012  & 1.5 & (0,0,-0.5) &
    (0,0,0)  & 16 & 0.01449 &  0.00007\\
    \verb|q3_0.5z| & 0031   & 3.0 & (0,0,0.5) & (0,0,0)& 14 &
    0.01734 &  $<10^{-4}$  \\
    \verb|q3_-0.5z| & 0038  & 3.0 & (0,0,-0.5) & (0,0,0) & 14 &
    0.01756 & $<10^{-4}$  \\
    \verb|q5_0.5z| & 0061   & 5.0 & (0,0,0.5) & (0,0,0) & 15 &
    0.01570 &  0.004  \\
    \verb|q5_-0.5z| & 0060   & 5.0 & (0,0,-0.5) & (0,0,0)& 15 &
    0.01591 &  0.003  \\
    \verb|q8_0.5z|& 0065  & 8.0 & (0,0,0.5) & (0,0,0) & 13 &
    0.01922 & 0.004\\
    \verb|q8_-0.5z| & 0064   & 8.0 & (0,0,-0.5) & (0,0,0) & 13 &
    0.01954 & 0.0005 \\
  \end{tabular}
  \end{ruledtabular}
\end{table*}

\subsubsection{Handling of spin terms}
\label{sec:PNOrders}

When constructing Taylor approximants that include the re-expansion of
the energy balance equation, the handling of spin terms becomes
important. In particular, terms of quadratic and higher order in
spins, such as $(\vec{S}_{i})^{2}$, appear in the evolution of
the orbital frequency at 3 PN and higher orders. These terms arise
from lower-order effects and represent incomplete information, since the
corresponding terms are unknown in the original power series for
the binding energy $E$ and the flux $\mathcal{F}$,

\begin{align}
\label{eq:EExpansion} 
E(x) &= -\frac{1}{2}\,m\nu x\,\left(1+\sum_{k=2}a_{k}x^{k/2}\right), \\
\mathcal{F}(x) &= \frac{32}{5}\nu^{2}x^{5}\left(1+\sum_{k=2}
  b_{k}x^{k/2}\right),
\end{align}
where $m=m_1+m_2$ and $\nu = m_1 m_2/m^2$, and
$m_{1,2}$ are the individual masses.

In these expansions, the spin-squared terms come in at 2 PN order and
thus appear in $a_4$ and $b_{4}$, cf. Eqs.~\eqref{eq:Energy-a4}
and~\eqref{eq:Flux-b4}.  Then, in the re-expansion series of Taylor
T4, 
\begin{equation}
  \label{eq:rexpand}
  S\equiv -\frac{\mathcal{F}}{dE/dx} = \frac{64 \nu}{5 m} x^5(1+\sum_{k=2}s_{k}x^{k/2}),
\end{equation}
 the coefficients $s_k$ can be recursively determined, e.g. 
\begin{align}
  \label{eq:rexpandsol}
  s_{4} &= b_{4}-3a_{4}-2s_{2}a_{2}, \\
 s_{6} &= b_{6}-(4a_{6}+3s_{2}a_{4}+\frac{5}{2}s_{3}a_{3}+2s_{4}a_{2}).
\end{align}
Thus, the spin-squared terms in $a_4$ and $b_4$ will induce
spin-squared terms at 3PN order in $s_{6}$.  The analogous conclusion
holds for Taylor T5.  These spin-squared terms are incomplete as the
corresponding terms in the binding energy and flux (i.e. in $a_6$ and
$b_6$) are not known.

This re-expansion has been handled in several ways in the
literature. For example, Nitz et~al.~\cite{Nitz:2013mxa} include only
terms which are linear in spin beyond 2 PN order.  On the other hand,
Santamar\'ia et~al.~\cite{Santamaria:2010yb} keep \emph{all} terms in
spin arising from known terms in $E$ and $\mathcal{F}$. In the present
work, we also keep all terms up to 3.5 PN order, which is the highest
order to which non-spin terms are completely known. Similarly, we
include all terms when computing the precession frequency (see
\ref{ap:orbitalEvolution}). We investigate the impact of different
spin-truncation choices in Sec.~\ref{sec:ChangeSpinTruncation}, along
with the impact of partially known 4 PN spin terms.

\subsection{Numerical Relativity Simulations}
\label{sec:NR}

To characterize the effectiveness of PN theory in reproducing NR
results, we have selected a subset of 16 simulations from the SXS
waveform catalog described in
Ref.~\cite{PhysRevLett.111.241104}.\footnote{The waveform and orbital
  data are publicly available at
  \url{https://www.black-holes.org/waveforms/}.} Our primary results
are based on six precessing simulations and a further ten
non-precessing ones for cross-comparisons.  To check for systematic
effects, we use a further 31 precessing simulations with random
mass-ratios and spins. The parameters of these runs are given in Table
\ref{tbl:Parameters}. They were chosen to represent various degrees of
complexity in the dynamics: (i) precessing versus non-precessing
simulations, the latter with spins parallel or anti-parallel to
$\ellHat$; (ii) one versus two spinning black holes; (iii) coverage of
mass ratio from $q=1$ to $q=8$; (iv) long simulations that cover more
than a precession cycle; and (v) a variety of orientations of
$\hat{\chi}_{1},\hat{\chi}_{2},\ellHat$.
Figure~\ref{fig:precessionCones} shows the precession cones of the
normal to the orbital plane and the spins for for the six primary
precessing cases in Table~\ref{tbl:Parameters}. The PN data were
computed using the Taylor T4 3.5 PN approximant.

The simulations from the catalog listed in Table~\ref{tbl:Parameters}
were run with numerical methods similar to~\cite{Buchman:2012dw}.  A
generalized harmonic evolution
system~\cite{Friedrich1985,Garfinkle2002,Pretorius2005c,Lindblom2006}
is employed, and the gauge is determined by gauge source functions
$H_a$.  During the inspiral phase of the simulations considered here,
$H_a$ is kept constant in the co-moving frame,
cf.~\cite{Scheel2009,Chu2009,Boyle2007}.  About 1.5 orbits before
merger, the gauge is changed to damped harmonic
gauge~\cite{Lindblom2009c,Szilagyi:2009qz,Choptuik:2009ww}.  This
gauge change happens outside the focus of the comparisons presented
here.

The simulation q5\_0.5x analyzed here is a re-run of the SXS
simulation SXS:BBH:0058 from Ref.~\cite{PhysRevLett.111.241104}.  We
performed this re-run for two reasons: First, SXS:BBH:0058 changes to
damped harmonic gauge in the middle of the inspiral, rather than close
to merger as all other cases considered in this work. Second,
SXS:BBH:0058 uses an unsatisfactorily low numerical resolution during
the calculation of the black hole spins.  Both these choices leave
noticeable imprints on the data from SXS:BBH:0058, and the re-run
q5\_0.5x allows us to quantify the impact of these deficiencies.  We
discuss these effects in detail in Secs.~\ref{sec:Gauge}
and~\ref{sec:AH-resolution}.  The re-run q5\_0.5x analyzed here is
performed with improved numerical techniques.  Most importantly,
damped harmonic gauge is used essentially from the start of the
simulation, $t\gtrsim 100M$.  The simulation q5\_0.5x also benefits
from improved adaptive mesh refinement~\cite{Szilagyi2014} and
improved methods for controlling the shape and size of the excision
boundaries; the latter methods are described in Sec.II.B. of
Ref.~\cite{Scheel:2014ina}.

We have performed convergence tests for some of the simulations;
Sec.~\ref{sec:tech_cons} will demonstrate with
Fig.~\ref{fig:convergenceWithLev} that numerical truncation error is
unimportant for the comparisons presented here.

\subsection{Characterizing Precession}
\label{CharacterizingPrecessionByRotors}

The symmetries of non-precessing systems greatly simplify the problem
of understanding the motion of the binary.  In a non-precessing
system, the spin vectors are essentially constant, and two of the
rotational degrees of freedom are eliminated in the binary's orbital
elements.  Assuming quasi-circular orbits, the entire system can be
described by the orbital phase $\Phi$, which can be defined as the
angle between $\nHat$ and $\hat{x}$.  In post-Newtonian theory the
separation between the black holes can be derived from $d\Phi/dt$.
Thus comparison between post-Newtonian and numerical orbits, for
example, reduces entirely to the comparison between $\Phi_{\text{PN}}$
and $\Phi_{\text{NR}}$~\cite{Buonanno-Cook-Pretorius:2007, Boyle2007}.
For precessing systems, on the other hand, the concept of an orbital
phase is insufficient; $\Phi$ could be thought of as just one of
the three Euler angles.  We saw in Sec.~\ref{sec:OrbitalEvolution}
that the orbital dynamics of a precessing system can be fairly
complex, involving the triad $(\nHat, \lambdaHat, \ellHat)$ (or
equivalently the frame rotor $\Rf$) as well as the two spin vectors
$\vec{\chi}_{1}$ and $\vec{\chi}_{2}$---each of which is, of course,
time dependent.  When comparing post-Newtonian and numerical results,
we need to measure differences between each of these quantities in
their respective systems.

To compare the positions and velocities of the black holes themselves,
we can condense the information about the triads into the quaternion
quantity~\cite{Boyle:2013}
\begin{equation}
  R_{\Delta} \defined \RfPN\, \RbarfNR~,
\end{equation}
which represents the rotation needed to align the PN frame with the NR
frame.  This is a geometrically meaningful measure of the relative
difference between two frames.  We can reduce this to a single real
number by taking the magnitude of the logarithm of this quantity,
defining the angle\footnote{More explanation of these expressions,
  along with relevant formulas for calculating their values, can be
  found in Appendix~\ref{sec:UsefulQuaternionFormulas}.}
\begin{equation}
  \label{eq:PhaseDifference}
  \Phi_{\Delta} \defined 2 \left\lvert \log R_{\Delta} \right\rvert~.
\end{equation}
This measure has various useful qualities.  It is invariant, in
  the sense that any basis frame used to define $\RfPN$ and $\RfNR$
  will result in the same value of $\Phi_{\Delta}$.  It conveniently
  distills the information about the difference between the frames
  into a single value, but is also non-degenerate in the sense that
  $\Phi_{\Delta} = 0$ if and only if the frames are identical.  It
  also reduces precisely to $\Phi_{\text{PN}} - \Phi_{\text{NR}}$ for
  non-precessing systems; for precessing systems it also incorporates
  contributions from the relative orientations of the orbital
  planes.%
\footnote{It is interesting to note that any attempt to define
  the orbital phases of precessing systems separately, and then
  compare them as some $\Phi_{B} - \Phi_{A}$, is either ill defined or
  degenerate---as shown in
  Appendix~\ref{sec:InadequacyOfSeparatePhases}.  This does not
    mean that it is impossible to define such phases, but at best they
    will be degenerate; multiple angles would be needed to represent
    the full dynamics.}

Despite these useful features of $\Phi_{\Delta}$, it may sometimes be
interesting to use different measures, to extract individual
components of the binary evolution.  For example,
Eq.~\eqref{eq:ellHatEv} describes the precession of the orbital plane.
When comparing this precession for two approaches, a more informative
quantity than $\Phi_{\Delta}$ is simply the angle between the
$\ellHat$ vectors in the two systems:
\begin{equation}
  \label{eq:LAngle}
  \angle L = \cos^{-1}\left(\ellHat^{\rm PN}\cdot\ellHat^{\rm NR}\right).
\end{equation}
Similarly, we will be interested in understanding the evolution of the
spin vectors, as given in Eqs.~\eqref{eq:SpinsEv}.  For this purpose,
we define the angles between the spin vectors:
\begin{subequations}
  \label{eq:ChiAngles}
  \begin{align}
    \angle \chi_{1} &= \cos^{-1}\left(\hat{\chi}_{1}^{\rm PN}\cdot\hat{\chi}_{1}^{\rm NR}\right), \\
    \angle \chi_{2} &=
                      \cos^{-1}\left(\hat{\chi}_{2}^{\rm PN}\cdot\hat{\chi}_{2}^{\rm NR}\right).
  \end{align}
\end{subequations}
We will use all four of these angles below to compare the
post-Newtonian and numerical orbital elements.

\begin{figure}
  \includegraphics[width=0.96\linewidth]{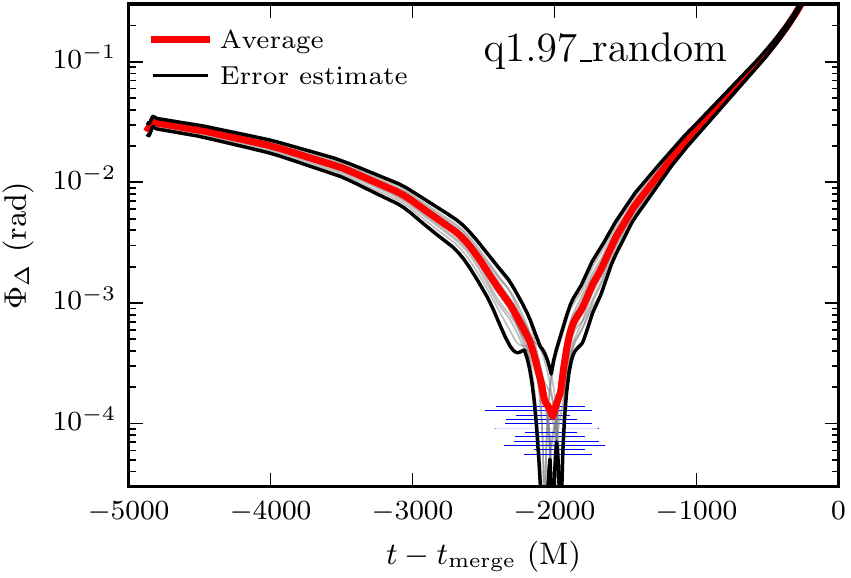} \\[10pt]
  \includegraphics[width=0.96\linewidth]{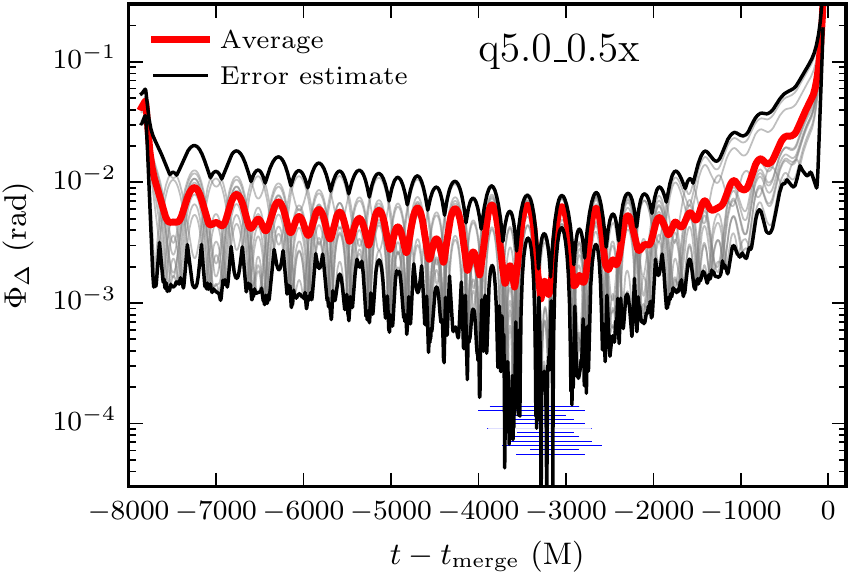}
  \caption{Examples of the averaging procedure and error estimates
    employed for all comparisons. Shown here are q1.97\_random and
    q5.0\_0.5x. PN evolutions were performed with the Taylor T1
    approximant. The thin blue lines show all the PN-NR matching
    intervals. 
  \label{fig:avProcedure}
}
\end{figure}

\subsection{Matching Post-Newtonian to Numerical Relativity}
\label{sec:matching}

When comparing PN theory to NR results, it is important to ensure that
the initial conditions used in both cases represent the same physical
situation. We choose a particular orbital frequency $\Omega_{m}$ and
use the NR data to convert it to a time $t_{m}$.  To initialize a PN
evolution at $t_{m}$, we need to specify
\begin{gather}
q,\chi_{1},\chi_{2}, \label{eq:conserved}\\
\ellHat,\nHat,\hat{\chi}_{1},\hat{\chi}_{2}, \label{eq:orientation} \\
\Omega. \label{eq:sep}
\end{gather}
The quantities \eqref{eq:conserved} are conserved during the PN
evolution. The quantities \eqref{eq:orientation} determine the
orientation of the the binary and its spins relative to the inertial
triad $(\hat{x},\hat{y},\hat{z})$. The orbital frequency $\Omega$ in
Eq.~\eqref{eq:sep}, finally, parametrizes the separation of the binary
at $t_{m}$. The simplest approach is to initialize the PN evolution
from the respective quantities in the initial data of the NR
evolution. This would neglect initial transients in NR data as in,
e.g., Fig.~1 of Ref.~\cite{Chu2009}. These transients affect the
masses and spins of the black holes, so any further PN-NR comparisons
would be comparing slightly different physical configurations. The NR
transients decay away within the first orbit of the NR simulation, so
one can consider initializing the PN evolution from NR at a time after
the NR run has settled down. However, the generally non-zero (albeit
very small) orbital eccentricity in the NR simulation can lead to
systematic errors in the subsequent comparison as pointed out in
Ref.~\cite{Boyle2007}.

Therefore, we use time-averaged quantities evaluated after the initial
transients have vanished. In particular, given a numerical relativity
simulation, we set the PN variables listed in Eq.~\eqref{eq:conserved}
to their numerical relativity values after junk radiation has
propagated away.

The remaining nine quantities Eqs.~\eqref{eq:orientation}
and~\eqref{eq:sep} must satisfy the constraint $\ellHat \cdot
\nHat\equiv 0$.  We determine them with constrained minimization by
first choosing an orbital frequency interval $[\Omega_m
-\delta\Omega/2, \Omega_m+\delta\Omega/2]$ of width $\delta\Omega$.
Computing the corresponding time interval $[t_{i},t_{f}]$ in the NR
simulation, we define the time average of any quantity $Q$ by
\begin{equation}
\langle Q\rangle=\frac{1}{t_f-t_i}\;\int_{t_i}^{t_f} Q\, dt.
\end{equation}
Using these averages, we construct the objective functional $\cal S$
as

\begin{equation}
\label{eq:ObjectiveFunction}
{\cal S}=\langle(\angle L)^{2}\rangle+\langle(\angle
\chi_{1})^{2}\rangle+\langle(\angle \chi_{2})^{2}\rangle + \langle(\Delta\Omega)^{2}\rangle
\end{equation}
where $\Delta\Omega = (\Omega_{\rm PN}-\Omega_{\rm NR})/\Omega_{\rm
  NR}$.  When a spin on the black holes is below $10^{-5}$ the
corresponding term is dropped from
Eq.~\eqref{eq:ObjectiveFunction}. The objective functional is then
minimized using the SLSQP algorithm~\cite{Kraft:1988, pyopt-paper} to
allow for constrained minimization.  In
Eq.~(\ref{eq:ObjectiveFunction}) we use equal weights for each term;
other choices of the weights do not change the qualitative picture
that we present.

The frequency interval $[\Omega_m\pm \delta\Omega/2]$ is chosen based
on several considerations. First it is selected after junk radiation
has propagated away. Secondly, it is made wide enough so that any
residual eccentricity effects average out.  Finally, we would like to
match PN and NR as early as possible.  But since we want to compare
various cases to each other, the lowest possible matching frequency
will be limited by the shortest NR run (case q8\_-0.5z).  Within these
constraints, we choose several matching intervals, in order to
estimate the impact of the choice of matching interval on our eventual
results.  Specifically, we use three matching frequencies
\begin{equation}
\label{eq:Omega_m}
m\Omega_m\in \{ 0.021067,0.021264,0.021461 \},
\end{equation}
and employ four different matching windows for each, namely
\begin{equation}
\label{eq:deltaOmega}
\delta\Omega/\Omega_m\in \{0.06,0.08,0.1,0.12\}.
\end{equation}
These frequencies correspond approximately to the range between 10-27
orbits to merger depending on the parameters of the binary, with the
lower limit for the case q1.0\_-0.5x and the upper for q8.0\_0.5x.

Matching at multiple frequencies and frequency windows allows an
estimate on the error in the matching and also ensures that the
results are not sensitive to the matching interval being used.  In
this article, we generally report results that are averaged over the
12 PN-NR comparisons performed with the different matching intervals.
We report error bars ranging from the smallest to the largest result
among the 12 matching intervals.  As examples,
Fig.~\ref{fig:avProcedure} shows $\Phi_{\Delta}$ as a function of time
to merger $t_{\rm merge}$ for the cases q1.97\_random and q5\_0.5x for
all the matching frequencies and intervals, as well as the average
result and an estimate of the error. Here $t_{\rm merge}$ is the time
in the NR simulation when the common horizon is detected.

\section{Results}
\label{sec:comparison}

\subsection{Precession Comparisons}
\label{sec:PrecessionComparisons}

We apply the matching procedure of Sec.~\ref{sec:matching} to the
precessing NR simulations in Table \ref{tbl:Parameters}. PN--NR
matching is always performed at the frequencies given by
Eq.~\eqref{eq:Omega_m} which are the lowest feasible orbital
frequencies across all cases in Table \ref{tbl:Parameters}.
Figure~\ref{fig:precessionCones} shows the precession cones for the
normal to the orbital plane $\ellHat$ and the spins
$\hat{\chi}_{1,2}$.   As
time progresses, $\ellHat$ and $\hat{\chi}_{1,2}$ undergo precession
and nutation, and the precession cone widens due to the emission of
gravitational radiation. Qualitatively, the PN results seem to follow
the NR results well, until close to merger.

\begin{figure}
  \includegraphics[width=0.98\linewidth]{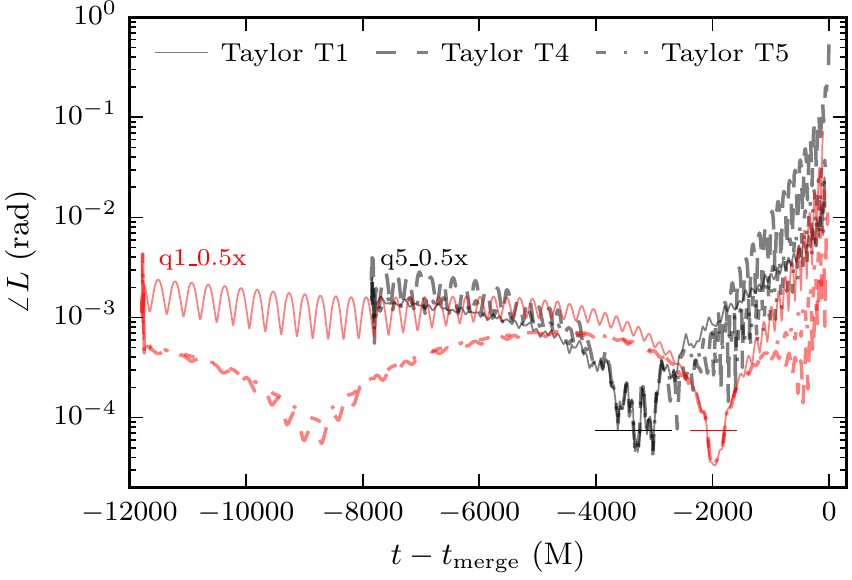}
  \caption{\label{fig:2casesLAngles}
    Angle $\angle L$ by which $\ellHat^{\rm PN}(t)$ differs from
    $\ellHat^{\rm NR}(t)$ for the configuration q1\_0.5x (red lines) and
    q5\_0.5x (black lines). $\angle L \le 0.2^{\circ}$ except very
    close to merger. In each case, the PN predictions based on
    different PN approximants are shown in different line styles. 
    Shown is the point-wise average of 12 $\angle L(t)$ curves, i.e. the
    thick red line of Fig.~\ref{fig:avProcedure}. The thin horizontal
    lines show the widest edges of the PN matching intervals.}
\end{figure}

We now turn to a quantitative analysis of the precession dynamics,
establishing first that the choice of Taylor approximant is of minor
importance for the precession dynamics.  We match PN dynamics to the
NR simulations q5\_0.5x and q1\_0.5x for the Taylor approximants T1,
T4 and T5.  We then compute the angles $\angle L$ and $\angle
\chi_{1}$.  Figure~\ref{fig:2casesLAngles} shows the resulting $\angle
L$.  During most of the inspiral, we find $\angle L$ of order a few
$10^{-3}$ radians increasing to $\sim 0.1$ radians during the last
$1000M$ before merger.  Thus the direction of the normal to the
orbital plane is reproduced well by PN theory. This result is
virtually independent of the Taylor approximant suggesting that the
choice of approximant only weakly influences how well PN precession
equations track the motion of the orbital plane.  In other words,
precession dynamics does not depend on details of orbital phasing like
the unmodeled higher-order terms in which the Taylor approximants
differ from each other.

\begin{figure}
  \includegraphics[width=0.98\linewidth]{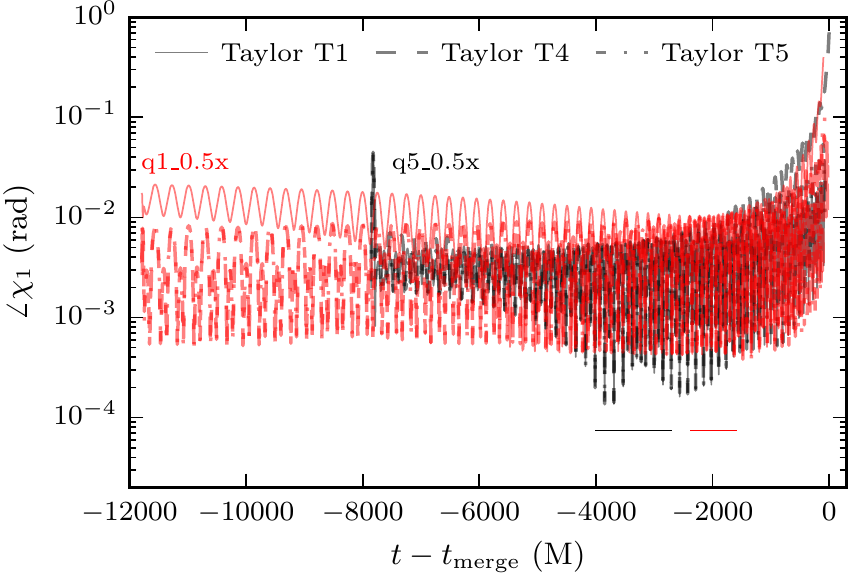}
  \caption{Angle $\angle \chi_{1}$ by which $\vec{\chi}_{1}^{\rm PN}(t)$
    differs from $\vec{\chi}_{1}^{\rm NR}(t)$ for the configuration
    q1\_0.5x (red lines) and q5\_0.5x (black lines).  In each case,
    the PN predictions based on different PN approximants are shown in
    different line styles. The thin horizontal lines show the widest
    edges of the PN matching intervals.}
  \label{fig:2casesSpinAngles}
\end{figure}

Turning to the spin direction $\hat{\chi}_{1}$ we compute the angle
$\angle \chi_{1}$ between $\hat{\chi}_{1}^{\rm NR}(t)$ and
$\hat\chi_{1}^{\rm PN}(t) $ and plot the result in
Fig.~\ref{fig:2casesSpinAngles}. While Fig.~\ref{fig:2casesSpinAngles}
looks busy, the first conclusion is that $\angle \chi_{1}$ is quite
small $\lesssim 0.01$ rad through most of the inspiral, and rises
somewhat close to merger.

The pronounced short-period oscillations of $\angle\chi_1$ in
Fig.~\ref{fig:2casesSpinAngles} are caused by differences between
PN-nutation features and NR-nutation features. To better understand
the nutation features and their impact on the angle $\angle \chi_{1}$,
we remove nutation features by filtering out all frequencies
comparable to the orbital frequency. This is possible because the
precession frequency is much smaller than the nutation frequency. The
filtering is performed with a 3rd order, bi-directional low pass
Butterworth filter~\cite{paarmann2001design} with a fixed cutoff
frequency chosen to be lower than the nutation frequency at the start
of the inspiral. Due to the nature of the filtering, the resulting
averaged spin will suffer from edge effects which affect approximately
the first and last 1000 M of the inspiral. Furthermore, the precession
frequency close to merger becomes comparable to the nutation frequency
at the start of the simulation and thus filtering is no longer
truthful in this region. Therefore, we only use the ``averaged'' spins
where such features are absent.

\begin{figure}
  \includegraphics[width=0.98\linewidth]{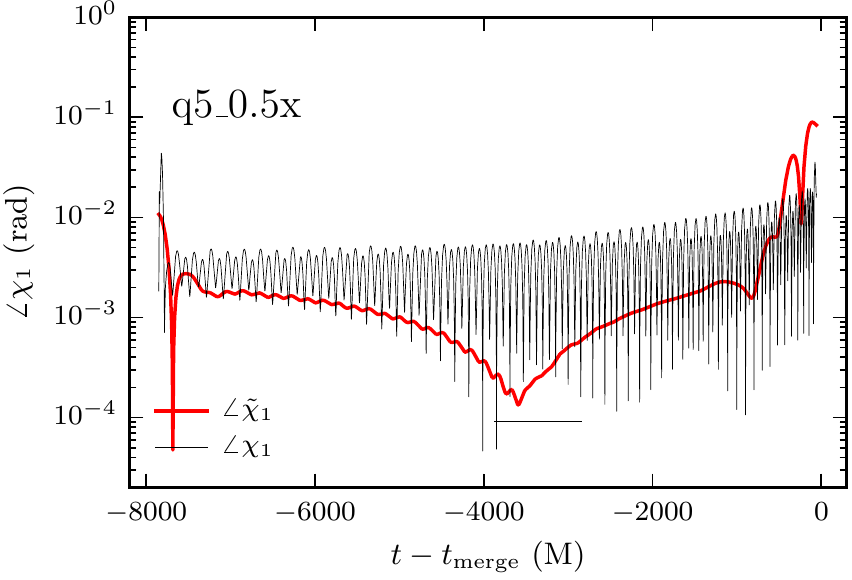}
  \caption{ \label{fig:AngleSmoothedS2cases} Angle $\angle
    \tilde{\chi}_{1}$ between the ``orbit-averaged'' spins for the
    configuration q5\_0.5x.  The non orbit-averaged difference
    $\angle\chi_{1}$ (cf. Fig.~\ref{fig:2casesSpinAngles}) is shown
    for comparison.  Shown is one matching interval as indicated by
    the thin horizontal line.  }
\end{figure}

Applying this smoothing procedure to both $\hat\chi_1^{\rm PN}$ and
$\hat\chi_1^{\rm NR}$ for the run q5\_0.5x, we compute the angle
$\angle\tilde\chi_1$ between the averaged spin vectors,
$\tilde\chi_1^{\rm PN}$ and $\tilde\chi_1^{\rm NR}$.  This angle is
plotted in Fig.~\ref{fig:AngleSmoothedS2cases}\footnote{To illustrate
  edge effects of the Butterworth filter,
  Fig.~\ref{fig:AngleSmoothedS2cases} includes the early and late time
  periods where the filter affects $\angle\tilde\chi_1$.}, where
results only for the Taylor T1 approximant are shown, and for only one
matching interval specified by $m\Omega_m=0.0210597$ and
$\delta\Omega/\Omega_m=0.1$. The orbit-averaged spin directions
$\tilde\chi_1^{\rm NR/PN}$ agree significantly better with each other
than the non-averaged ones (cf. the black line in
Fig.~\ref{fig:AngleSmoothedS2cases}, which is duplicated from
Fig.~\ref{fig:2casesSpinAngles}).  In fact, the orbit-averaged spin
precessing between NR and PN agrees as well as the orbital angular
momentum precession, cf. Fig.~\ref{fig:2casesLAngles}.  Thus, the
difference in the spin dynamics is dominated by the nutation features,
with the orbit-averaged spin dynamics agreeing well between PN and NR.

\begin{figure}
  \includegraphics[width=0.98\linewidth]{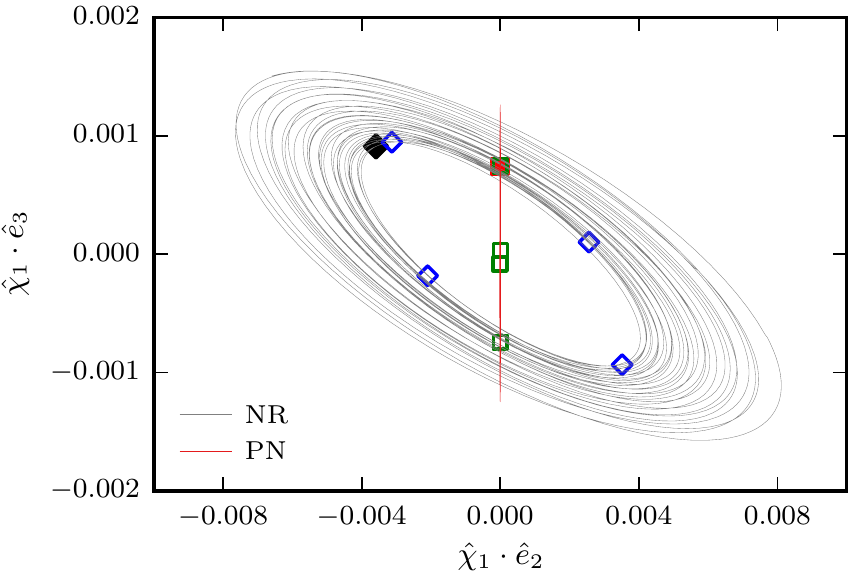}
  \caption{ \label{fig:ProjectionSpinsQ5} The projection of
    $\hat{\chi}_{1}^{\rm NR}$ and $\hat{\chi}_{1}^{\rm PN}$ onto the
    $\hat{e}_{2}-\hat{e}_{3}$ plane described in the text for case
    q5\_0.5x.  The system is shown in the interval $t-t_{\rm
      merge}\in[-6662,-1556]$.  along the $\hat{e}_{3}$
    axis. Meanwhile, the NR data show variations in $\hat{e}_{2}$ and
    $\hat{e}_{3}$ directions of comparable magnitude. The solid
    symbols (black diamond for NR, red square for PN) indicate the
    data at the start of the plotted interval, chosen such that
    $\hat\chi_1\cdot\hat n$ is maximal---i.e., where the spin
    projection into the orbital plane is parallel to $\hat n$.  The
    subsequent four open symbols (blue diamonds for NR, green squares
    for PN) indicating the position 1/8-th, 1/4-th, 3/8-th and 1/2 of
    an orbit later. }
\end{figure}

\begin{figure}
  \includegraphics[width=0.96\linewidth]{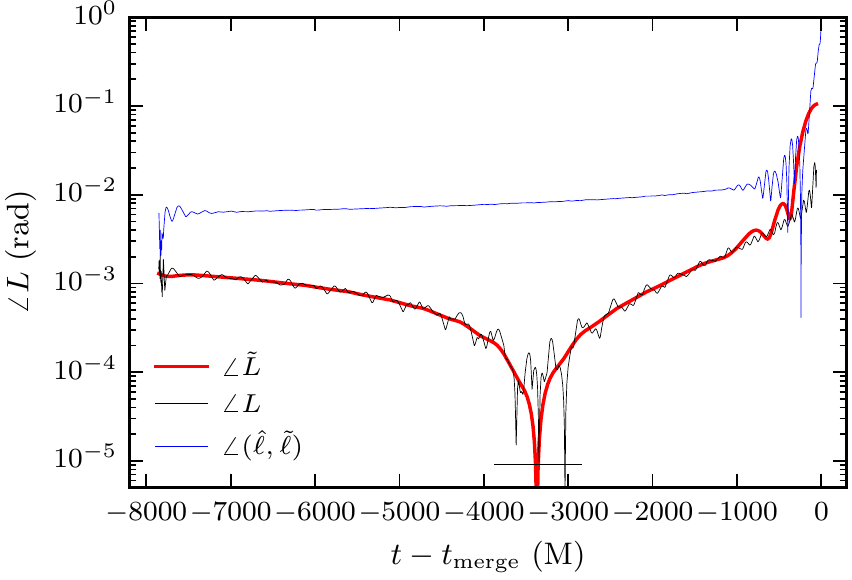} \\[10pt]
  \includegraphics[width=\linewidth]{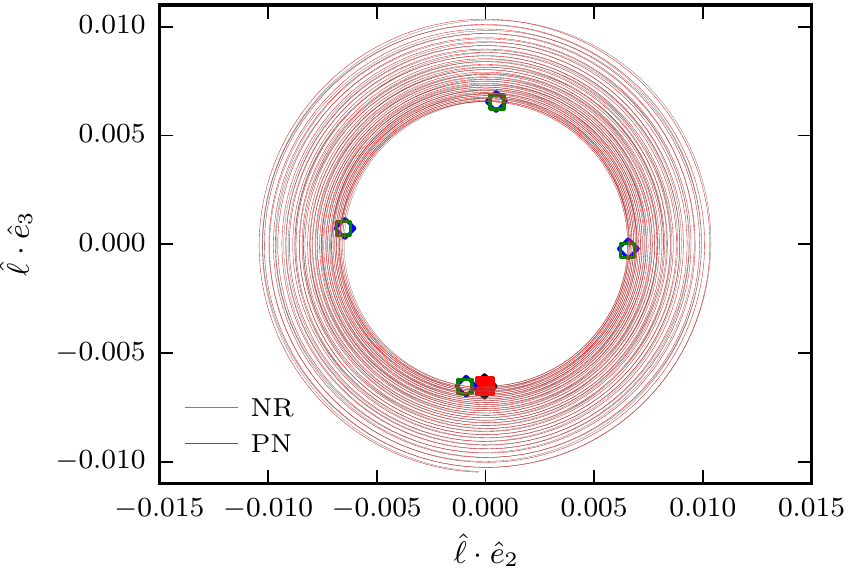}
  \caption{ \label{fig:ProjectionEllHatQ5} Characterization of
    nutation effects of the orbital angular momentum.  {\bf Top}:
    angle $\angle \tilde{L}$ between the ``averaged'' $\ellHat$ in PN
    and NR for the configuration q5\_0.5x (thick red line).  $\angle
    L$ is shown in thin black line for comparison (cf.
    Fig.~\ref{fig:AngleSmoothedS2cases}). The thin blue line shows
    $\angle (\ellHat,\tilde{\ell})$ between the averaged and the
    filtered signal. Note that it is larger than both $\angle L$ and
    $\angle \tilde{L}$.  {\bf Bottom}: the projection of $\ellHat^{\rm
      NR}$ (gray) and $\ellHat^{\rm PN}$ (red) onto the
    $\hat{e}_{2}-\hat{e}_{3}$ plane described in the text for case
    q5\_0.5x (cf. Fig.~\ref{fig:ProjectionEllHatQ5}).  The system is
    shown in the interval $[-6662,-1556]$. Both PN and NR show the
    same behavior, in contrast to the behavior of the spin in
    Fig.~\ref{fig:ProjectionSpinsQ5}. The PN-NR matching interval is
    indicated by the horizontal line in the top panel. }
\end{figure}

To characterize the nutation features in the spin vectors, we
introduce a coordinate system which is specially adapted to
highlighting nutation effects.  The idea is to visualize nutation with
respect to the averaged spin vector $\tilde\chi$.  We compute the
time-derivative $\dot{\tilde{\chi}}$ numerically.  Assuming that the
``averaged'' spin is undergoing pure precession, so that
$\tilde{\chi}\cdot\dot{\tilde{\chi}} = 0$, we define a new coordinate
system $(\hat{e}_{1},\hat{e}_{2},\hat{e}_{3})$ by
$\hat{e}_{1}=\tilde{\chi},\hat{e}_{2}=\dot{\tilde{\chi}}/|\dot{\tilde{\chi}}|,
\hat{e}_{3} = \hat{e}_1\times \hat{e}_{2} $.  The spin is now
projected onto the $\hat{e}_{2}-\hat{e}_{3}$ plane, thus showing the
motion of the spin in a frame ``coprecessing'' with the averaged
spin. This allows us to approximately decouple precession and nutation
and compare them separately between PN and NR.

Figure~\ref{fig:ProjectionSpinsQ5} plots the projection of the spins
$\chi_1^{\rm NR}$ and $\chi_1^{\rm PN}$ onto their respective ``orbit
averaged'' $\hat{e}_{2}-\hat{e}_{3}$ planes.  We see that the behavior
of the NR spin and the PN spins are qualitatively different: For this
single-spin system, the PN spin essentially changes only in the
$\hat{e}_3$ direction (i.e., orthogonal to its average motion
$\dot{\tilde{\chi}}^{\rm PN}$).  In contrast, the NR spin undergoes
elliptical motion with the excursion along its $\hat e_2$ axis (i.e.,
along the direction of the average motion) about several times larger
than the oscillations along $\hat e_{3}$.  The symbols plotted in
Fig.~\ref{fig:ProjectionSpinsQ5} reveal that each of the elliptic
``orbits'' corresponds approximately to half an orbit of the binary,
consistent with the interpretation of this motion as nutation.  The
features exhibited in Fig.~\ref{fig:ProjectionEllHatQ5} are similar
across all the single-spinning precessing cases considered in this
work.  The small variations in spin direction exhibited in
Fig.~\ref{fig:ProjectionSpinsQ5} are orders of magnitude smaller than
parameter estimation capabilities of LIGO, e.g.~\cite{Veitch:2014wba},
and so we do not expect that these nutation features will have a
negative impact on GW detectors.

\begin{figure*}
  \includegraphics[width=0.48\linewidth]{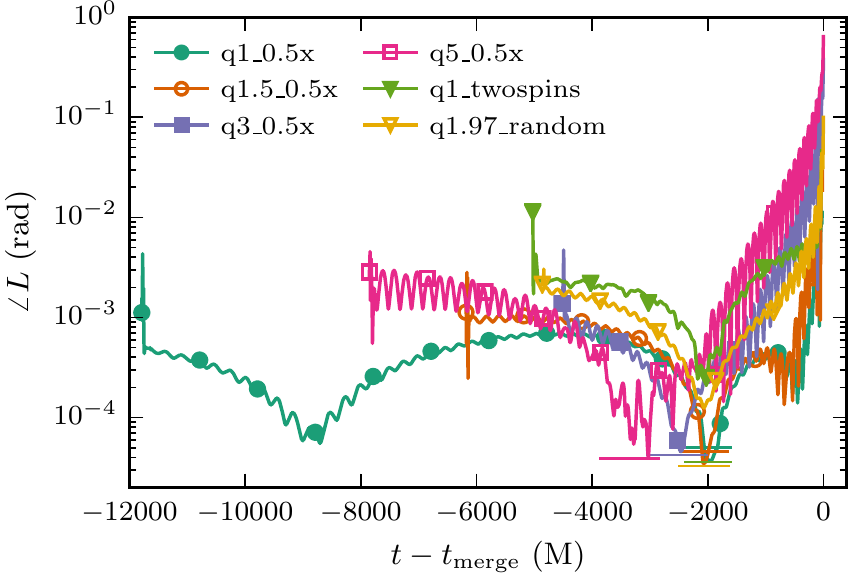} \hfill
  \includegraphics[width=0.48\linewidth]{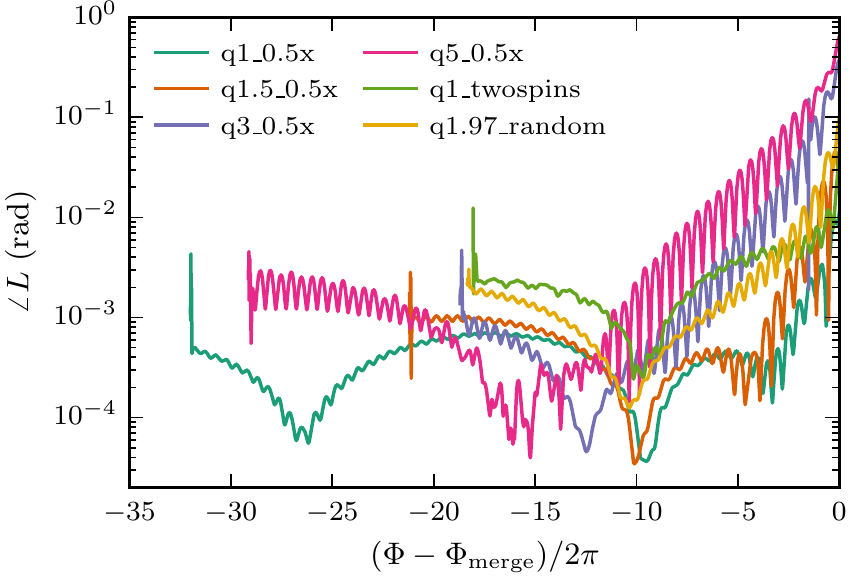} \\[0.2cm]
  \includegraphics[width=0.48\linewidth]{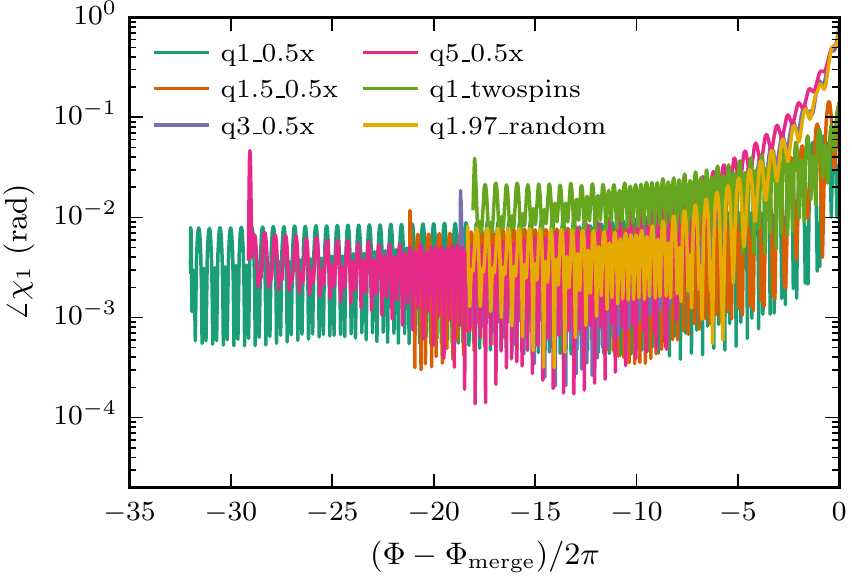} \hfill
  \includegraphics[width=0.48\linewidth]{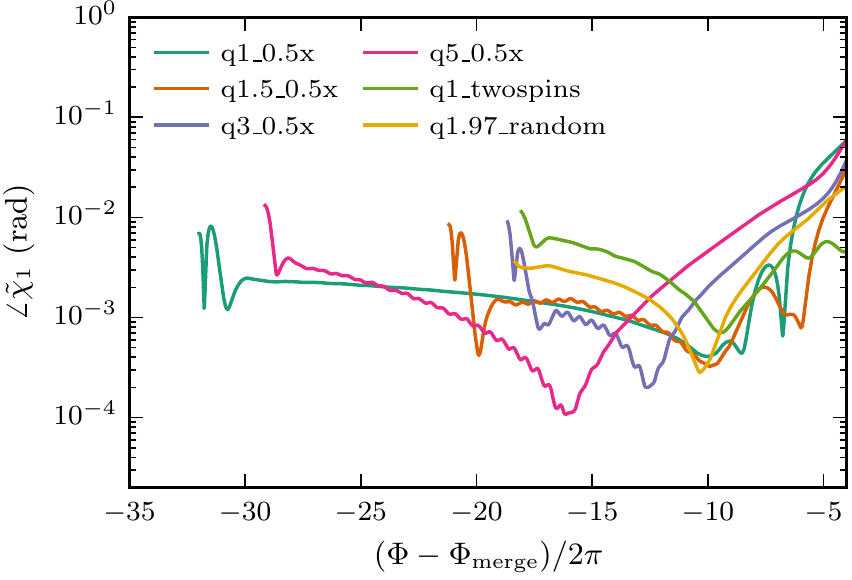}
  \caption{ \label{fig:AngleLAll} Comparison of orbital plane and spin
    precession for the primary six precessing NR simulations.  {\bf
      Top Left}: $\angle L$ as a function of time to merger.  {\bf Top
      right}: $\angle L$ as a function of \emph{orbital phase}.  {\bf
      Bottom left}: $\angle \chi_{1}$ as a function of orbital
    phase. {\bf Bottom right}: $\angle \tilde{\chi}_{1}$ between the
    averaged spins.  All data plotted are averages over 12 matching
    intervals, cf. Fig.~\ref{fig:avProcedure}, utilizing the Taylor T4
    PN approximant. The thin horizontal lines in the top left panel
    show the widest edges of the PN matching intervals.}
\end{figure*}

Let us now apply our nutation analysis to the orbital angular momentum
directions $\ellHat$.  Analogous to the spin, we compute averages
$\tilde\ell^{\rm NR}$ and $\tilde\ell^{\rm PN}$, and compute the angle
between the directions of the averages, $\angle\tilde
L=\angle\left(\tilde\ell^{\rm PN}, \tilde\ell^{\rm NR}\right)$.  This
angle---plotted in the top panel of
Fig.~\ref{fig:ProjectionEllHatQ5}---agrees very well with the
difference $\angle L$ that was computed without orbit-averaging.  This
indicates that the nutation features of $\hat\ell$ agree between NR
and PN.  The top panel of Fig.~\ref{fig:AngleLAll} also plots the
angle between the raw $\hat\ell^{\rm NR}$ and the averaged
$\tilde\ell^{\rm NR}$, i.e. the opening angle of the nutation
oscillations.  As is apparent in Fig.~\ref{fig:ProjectionEllHatQ5},
the angle between $\hat\ell^{\rm NR}$ and $\tilde\ell^{\rm NR}$ is
about 10 times larger than the difference between NR and PN ($\angle
L$ or $\angle \tilde L$), confirming that
nutation features are captured.  The lower panel of
Fig.~\ref{fig:ProjectionEllHatQ5} shows the projection of $\hat\ell$
orthogonal to the direction of the average $\tilde\ell$.  In contrast
to the spins shown in Fig.~\ref{fig:ProjectionSpinsQ5}, the nutation
behavior of $\ellHat$ is in close agreement between NR and PN: For
both, $\ellHat$ precesses in a circle around $\tilde\ell$, with
identical period, phasing, and with almost identical amplitude.  We
also point out that the shape of the nutation features differs between
$\ellHat$ and $\hat{\chi}_{1}$: $\ellHat$ circles twice per orbit
around its average $\tilde\ell$, on an almost perfect circle with
equal amplitude in the $\hat e_2$ and $\hat e_3$ direction.

We now extend our precession dynamics analysis to the remaining five
primary precessing NR simulations listed in
Table~\ref{tbl:Parameters}. The top left panel of Figure
\ref{fig:AngleLAll} shows $\angle L$.  The difference in the direction
of the normal to the orbital plane is small; generally $\angle
L\lesssim 10^{-2}$ radians, except close to merger. Thus it is evident
that the trends seen in Fig.~\ref{fig:2casesLAngles} for $\angle L$
hold across all the precessing cases. To make this behavior clearer,
we parameterize the inspiral using the orbital phase instead of time,
by plotting the angles versus the orbital phase in the NR simulation,
as shown in the top right panel of Fig.~\ref{fig:AngleLAll}.
Thus, until a few orbits to merger PN represents the precession and
nutation of the orbital plane well.

The bottom left panel of Fig.~\ref{fig:AngleLAll} establishes
qualitatively good agreement for $\angle\chi_1$, with slightly higher
values than $\angle L$.  As already illustrated in
Fig.~\ref{fig:AngleSmoothedS2cases}, nutation features dominate the
difference.  Averaging away the nutation features, we plot the angle
$\angle \tilde{\chi}_{1}$ between the smoothed spins in the bottom
left panel of Fig.~\ref{fig:AngleLAll}, where the behavior of $\angle
\chi_{1}$ is very similar to that of $\angle L$.  This confirms that
the main disagreement between PN and NR spin dynamics comes from
nutation features, and suggests that the secular precession of the
spins is well captured across all cases, whereas the nutation of the
spins is not.

\begin{figure}
  \centering
  \includegraphics[width=0.96\linewidth]{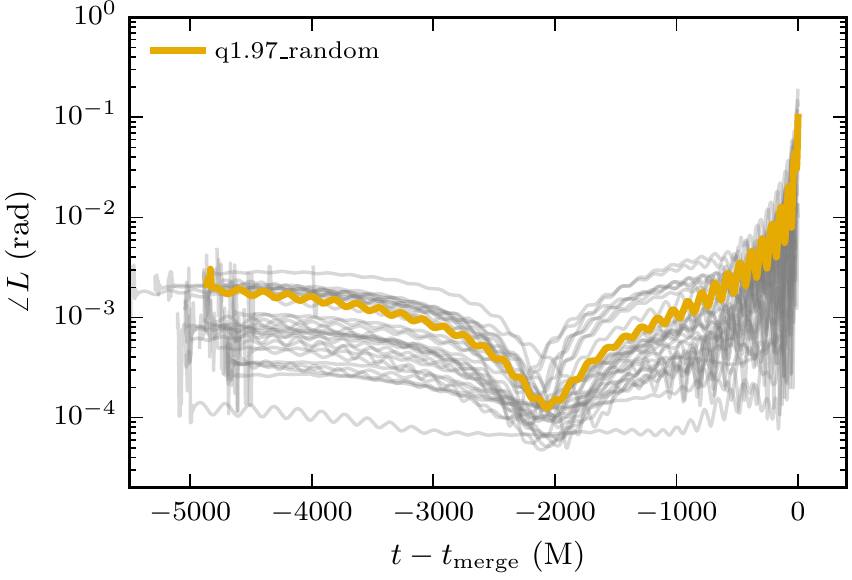}
  \caption{  \label{fig:AngleLRandom31}
    $\angle L$ for additional 31 precessing configurations with
    arbitrary oriented spins as well as the case q1.97\_random. Here
    $q\in(1,2),\ \chi_{1,2}\le 0.5$. For all cases, $\angle L <
    0.5^{\circ}$ throughout most of the inspiral.
  All data plotted
    are averages over 12 matching intervals,
    cf. Fig.~\ref{fig:avProcedure}.}
\end{figure}

All configurations considered so far except q1.97\_random have
$\vec{S}\cdot\ellHat=0$ at the start of the simulations, where $\vec
S=\vec S_1+\vec S_2$ is the total spin angular momentum of the
system.  When $\vec{S}\cdot\ellHat=0$, several terms in PN equations
vanish, in particular the spin orbit terms in the expansions of the
binding energy, the flux and the orbital precession frequency, see
Eqs.~\eqref{eq:EExpansion}, \eqref{eq:FExpansion},
and~\eqref{eq:varpi} in Appendix A.

To verify whether $\vec{S}\cdot\ellHat=0$ introduces a bias to our
analysis, we perform our comparison on an additional set of 31
binaries with randomly oriented spins. These binaries have mass ratio
$1\le q \le 2$, spin magnitudes $0\le\chi_{1,2}\le0.5$, and correspond
to cases SXS:BBH:0115 - SXS:BBH:0146 in the SXS
catalog. Fig.~\ref{fig:AngleLRandom31} plots $\angle L$ for these
additional 31 PN-NR comparisons in gray, with q1.97\_random
highlighted in orange. The disagreement between PN and NR is similarly
small in all of these cases, leading us to conclude that our results
are robust in this region of the parameter space.

\subsection{Orbital Phase Comparisons}
\label{sec:OrbitalPhaseComparisons}

\begin{figure}
  \centering
  \includegraphics[width=0.98\linewidth]{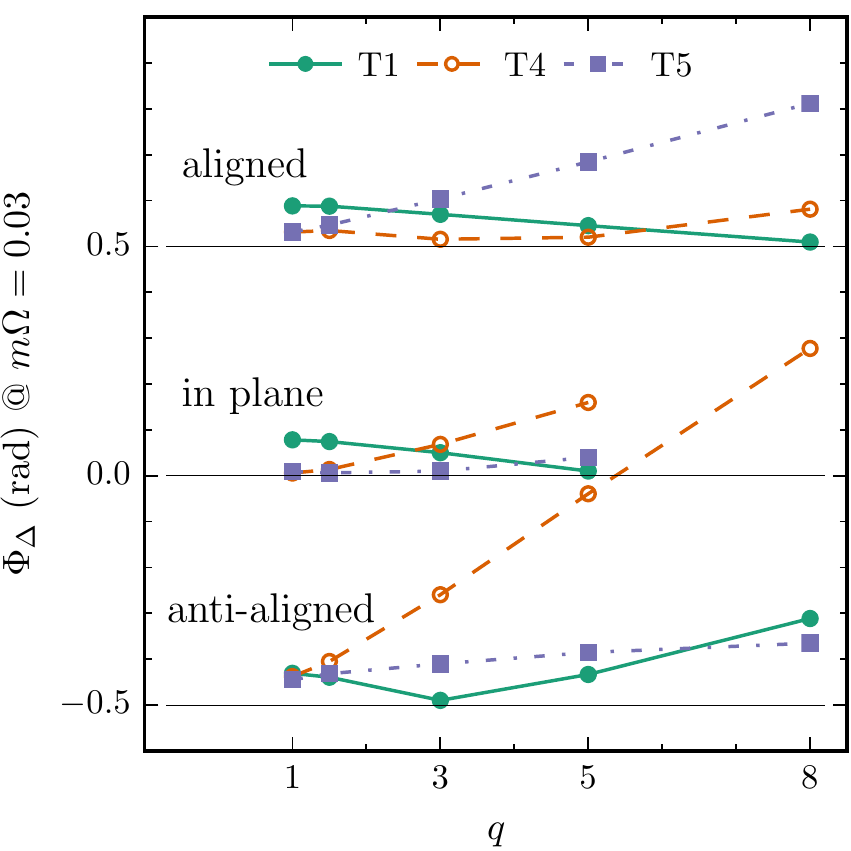}
  \caption{\label{fig:massRatioDep} $\Phi_{\Delta}$ as a function of
    mass ratio for BBH systems with $\chi_1=0.5$, and spin direction
    aligned (top), orthogonal (middle), and anti-aligned (bottom) with
    the orbital angular momentum.  For clarity, the
    aligned/anti-aligned data are offset by $+0.5$ and $-0.5$,
    respectively, with the thin horizontal black lines indicating zero
    for each set of curves.  Plotted is $\Phi_\Delta$ averaged over
    the 12 matching intervals, cf. Fig.~\ref{fig:avProcedure}, and for
    three different Taylor approximants.}
\end{figure}

Along with the precession quantities described above, the orbital
phase plays a key role in constructing PN waveforms.  We use
$\Phi_{\Delta}$, a geometrically invariant angle that reduces to the
orbital phase difference for non-precessing binaries
(cf. Sec.~\ref{CharacterizingPrecessionByRotors}) to characterize
phasing effects.  We focus on single spin systems with mass-ratios
from 1 to 8, where the more massive black hole carries a spin of
$\chi_1=0.5$, and where the spin is aligned or anti-aligned with the
orbital angular momentum, or where the spin is initially tangent to
the orbital plane.  We match all NR simulations to post-Newtonian
inspiral dynamics as described in Sec.~\ref{sec:matching}, using the
12 matching intervals specified in Eqs.~\eqref{eq:Omega_m}
and~\eqref{eq:deltaOmega}.  We then compute the phase difference
$\Phi_\Delta$ at the time at which the NR simulation reaches orbital
frequency $m\Omega=0.3$.

The results are presented in Fig.~\ref{fig:massRatioDep}, grouped
based on the initial orientation of the spins: aligned, anti-aligned,
and in the initial orbital plane. For aligned runs, there are clear
trends for Taylor T1 and T5 approximants: for T1, differences decrease
with increasing mass ratio (at least up to $q=8$); for T5, differences
increase.  For Taylor T4, the phase difference $\Phi_\Delta$ has a
minimum and there is an overall increase for higher mass ratios. For
anti-aligned runs, Taylor T5 shows the same trends as for the aligned
spins.  Taylor T4 and T1 behaviors, however, have reversed: T4
demonstrates a clear increasing trend with mass ratio, whereas T1
passes through a minimum with overall increases for higher mass
ratios. Our results are also qualitatively consistent with the results
described in \cite{HannamEtAl:2010} as we find that for equal mass
binaries, the Taylor T4 approximant performs better than the Taylor T1
approximant (both for aligned and anti-aligned spins).

For the in-plane precessing runs, we see clear trends for all 3
approximants: Taylor T4 and T5 both show increasing differences with
increasing mass ratio, and T1 shows decreasing differences. These
trends for precessing binaries are consistent with previous work on
non-spinning binaries~\cite{MacDonald:2012mp}, which is expected since
for $\vec{S}\cdot\ellHat$ many of the same terms in the binding energy
and flux vanish as for non-spinning binaries.  Overall, we find that
for different orientations and mass ratios, no one Taylor approximant
performs better than the rest, as expected if the differences between
the approximants arise from different treatment of higher-order terms.

\subsection{Convergence with PN order}
\label{sec:PNConvergence}

\begin{figure}
  \centering
  \includegraphics[width=0.96\linewidth, trim=0 26 0 0]{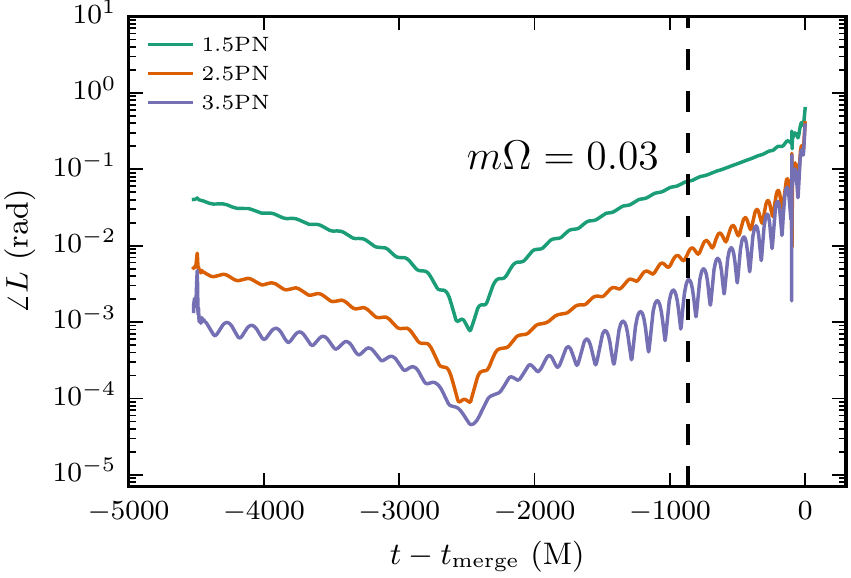}\\
  \includegraphics[width=0.96\linewidth]{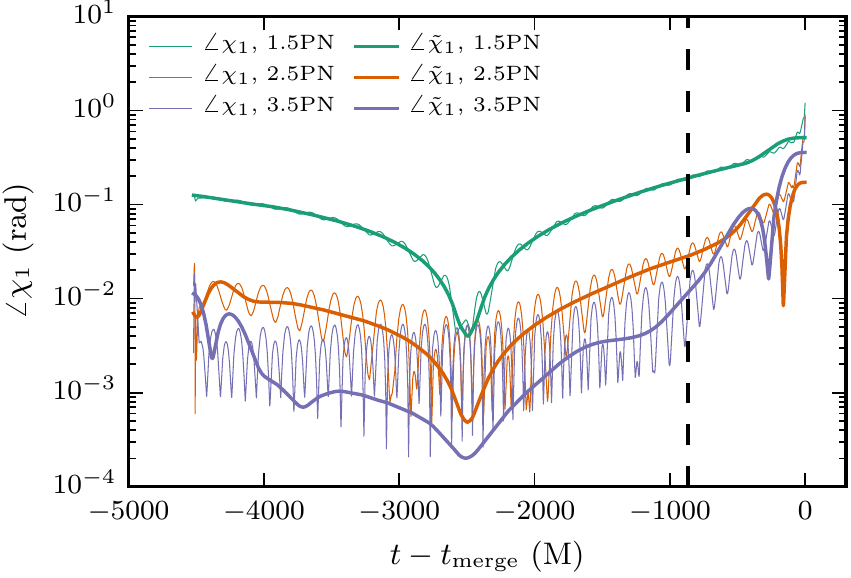}
  \caption{Comparison of PN-NR precession dynamics when the expansion
    order of the PN precession equations is varied.  Shown is the case
    q3\_0.5x. The top panel shows the precession of the orbital plane,
    and the bottom panel of the spin $\hat\chi_1$ (without and with
    averaging). All data shown are averages over 12 matching
    intervals, cf. Fig.~\ref{fig:avProcedure}.}
  \label{fig:q3PrecConv}
\end{figure}

So far all comparisons were performed using all available
post-Newtonian information.  It is also instructive to consider
behavior at different PN order, as this reveals the convergence
properties of the PN series, and allows estimates of how accurate
higher order PN expressions might be.

The precession frequency $\varpi$, given in Eq.~\eqref{eq:varpi}, is a
product of series in the frequency parameter $x$.  We multiply out
this product, and truncate it at various PN orders from leading order
(corresponding to 1.5PN) through next-to-next-to-leading order
(corresponding to 3.5PN).  Similarly, the spin precession frequencies
$\vec\Omega_{1,2}$ in Eqs.~\eqref{eq:SpinsEv}
and~\eqref{eq:OmegaSpinsEv} are power series in $x$.  We truncate the
power series for $\vec\Omega_{1,2}$ in the same fashion as the power
series for $\varpi$, but keep the orbital phase evolution at 3.5PN
order, where we use the TaylorT4 prescription to implement the energy
flux balance.  For different precession-truncation orders, we match
the PN dynamics to the NR simulations with the same techniques and at
the same matching frequencies as in the preceding sections.

When applied to the NR simulation q3\_0.5x, we obtain the results
shown in Fig.~\ref{fig:q3PrecConv}.  This figure shows clearly that
with increasing PN order in the precession equations, PN precession
dynamics tracks the NR simulation more and more accurately.  When only
the leading order terms of the precession equations are included
(1.5PN order), $\angle L$ and $\angle\chi_1$ are $\approx 0.1$rad; at
3.5PN order this difference drops by nearly two orders of magnitude.

\begin{figure}
  \centering
  \includegraphics[width=0.96\linewidth,trim=0 22 0 0,clip=true]{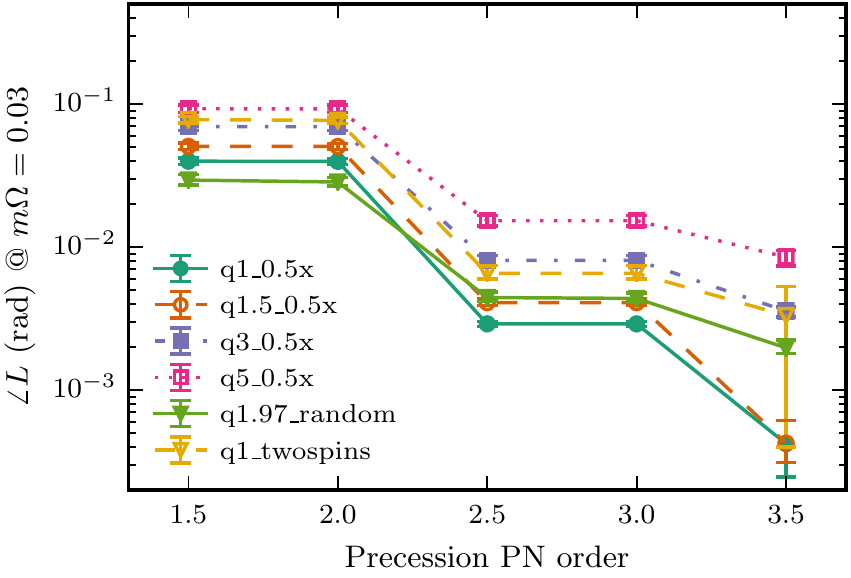}\\
  \includegraphics[width=0.96\linewidth]{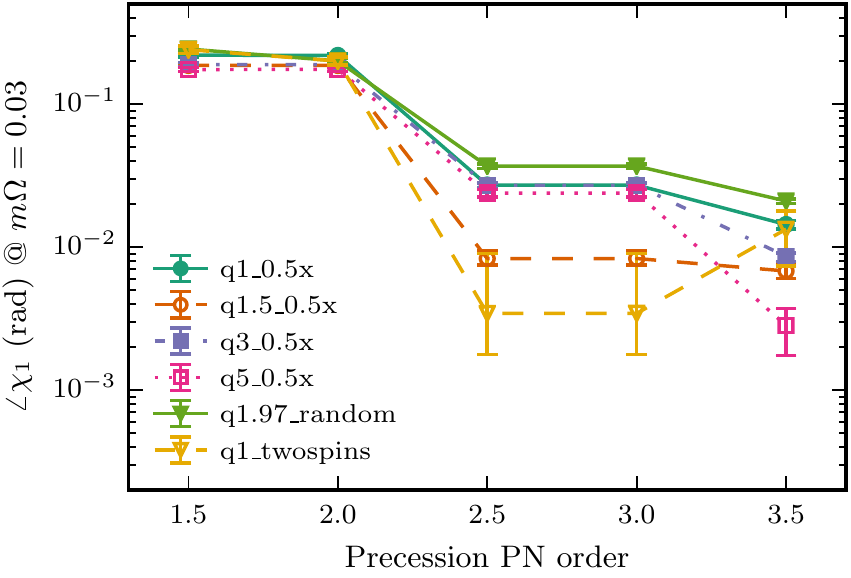}
  \caption{Convergence of the PN precession equations for all cases in
    Table \ref{tbl:Parameters}. The evolution was done with the Taylor
    T4 approximant at 3.5 PN order. The leading order spin-orbit
    correction is at 1.5 PN order and the spin-squared corrections
    appear at 2 PN order.  Each data point is the average $\angle L$
    over PN-NR comparisons performed using 12 matching intervals,
    cf. Fig.~\ref{fig:avProcedure}, with error bars showing the maximal
    and minimal $\angle L$ and $\angle \chi_1$ of the 12 fits.  }
  \label{fig:PrecPNOrder}
\end{figure}

We repeat this comparison for our six main precessing cases from
Table~\ref{tbl:Parameters}. The results are shown in
Fig.~\ref{fig:PrecPNOrder}. It is evident that for \emph{all cases}
$\angle L $ decreases with increasing order in the precession
equations with almost 2 orders of magnitude improvement between
leading order and next-to-next leading order truncations.  A similar
trend is seen in the convergence of the spin angle $\angle\chi_{1}$
shown in bottom panel of Fig.~\ref{fig:PrecPNOrder}. The angle
decreases with PN order almost monotonically for all cases except
q1.0\_twospins.  However, this is an artificial consequence of picking
a particular matching point at $m\Omega=0.03$: as can be seen from the
bottom panel of Fig.~\ref{fig:q3PrecConv} $\angle \chi_{1}$ shows
large oscillations and it is a coincidence that the matching point
happens to be in a ``trough'' of $\chi_{1}$.

So far we have varied the PN order of the precession equations, while
keeping the orbital frequency evolution at 3.5PN order.  Let us now
investigate the opposite case: varying the PN order of the orbital
frequency and monitoring its impact on the orbital phase evolution.
We keep the PN order of the precession equations at 3.5PN, and match
PN with different orders of the orbital frequency evolution (and
TaylorT4 energy-balance prescription) to the NR simulations.  We then
evaluate $\Phi_\Delta$ (a quantity that reduces to the orbital phase
difference in cases where the latter is unambiguously defined) at the
time at which the NR simulation reaches the frequency $m\Omega=0.03$.
We examine our six primary  precessing runs,
and also the aligned-spin and anti-aligned spin binaries listed in
Table~\ref{tbl:Parameters}.

\begin{figure}
  \includegraphics[width=0.96\linewidth,trim=0 26 0 0]{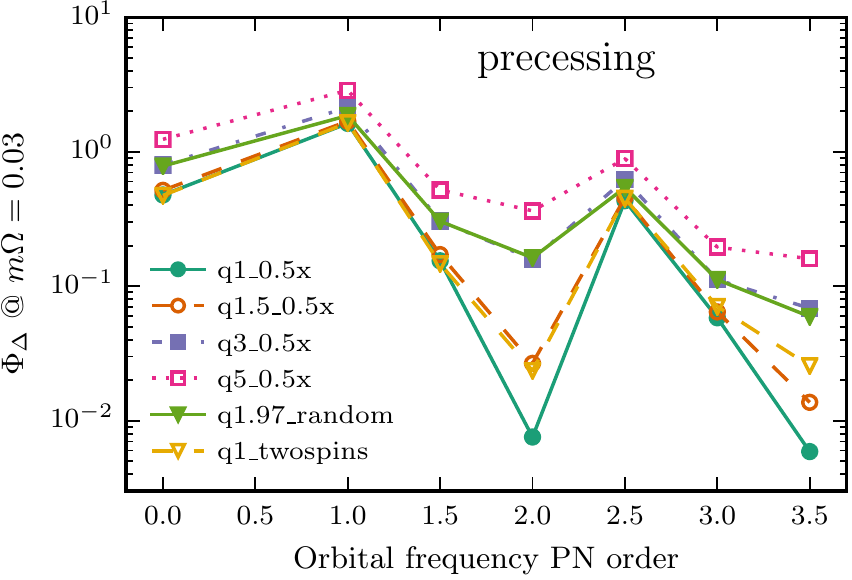}\\
  \includegraphics[width=0.96\linewidth,trim=0 26 0 0]{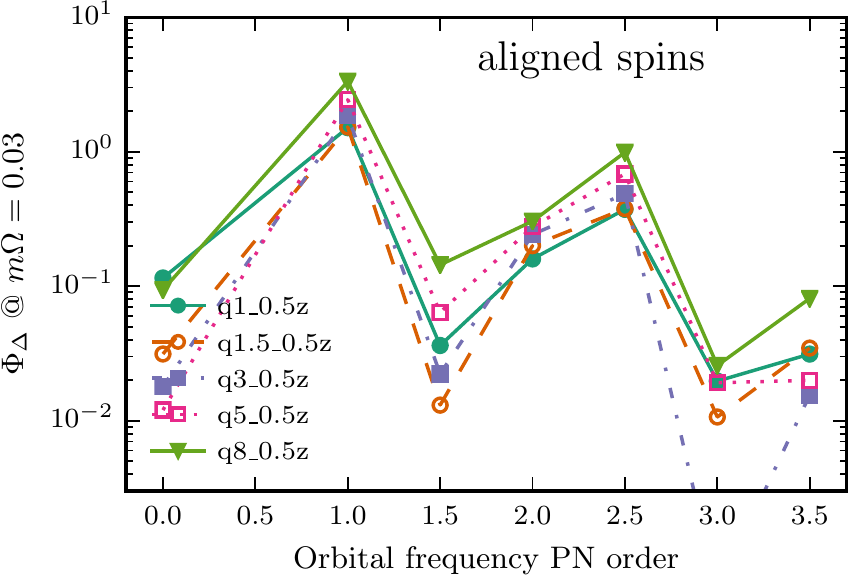}\\
   \includegraphics[width=0.96\linewidth]{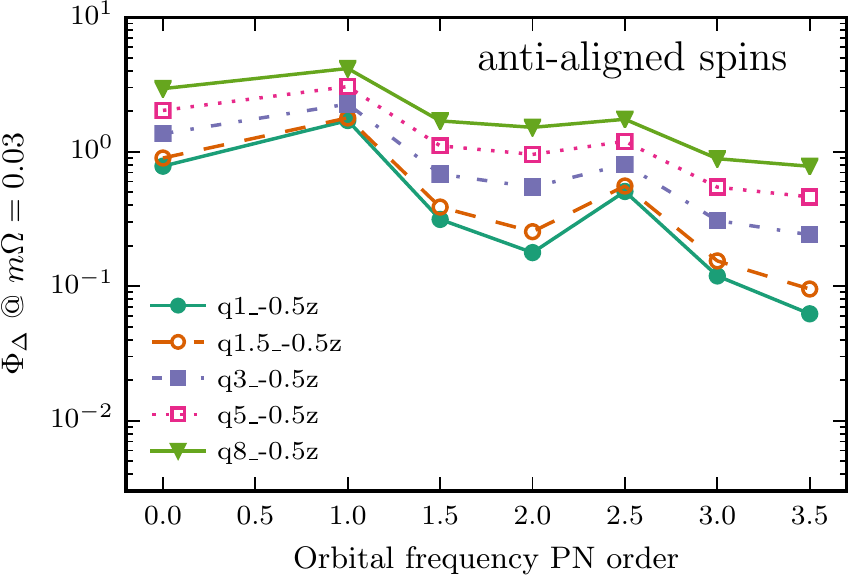}
   \caption{Convergence of the Taylor T4 approximant with PN
     order. Shown are all cases from Table \ref{tbl:Parameters}. {\bf
       Top}: all precessing cases. {\bf Middle}: aligned spin
     cases. {\bf Bottom}: anti-aligned spin cases.  Each data point
     shown is averaged over PN-NR comparison with 12 matching
     intervals, cf. Fig.~\ref{fig:avProcedure}.  Error bars are
     omitted for clarity, but would be of similar size to those in
     Fig.~\ref{fig:spinTruncationSummary}.}
  \label{fig:q3_OmegaConv}
\end{figure}

When the spin is initially in the orbital plane, as seen in the top
panel of Fig.~\ref{fig:q3_OmegaConv}, the overall trend is a
non-monotonic error decrease with PN order, with spikes at 1 and 2.5
PN orders as has been seen previously with non-spinning
binaries~\cite{Boyle2007}.  All of the aligned cases show a large
improvement at 1.5 PN order, associated with the leading order
spin-orbit contribution. The phase differences then spike at 2 and 2.5
PN orders and then decrease at 3 PN order. Finally, different cases
show different results at 3.5 PN with some showing decreases
differences while for others the differences increase.

For the anti-aligned cases the picture is similar to precessing cases
with a spike at 1 and 2.5 PN orders and monotonic improvement
thereafter. The main difference from precessing cases is the magnitude
of the phase differences, which is larger by a factor of $\sim 5$ at
3.5 PN order for the anti-aligned cases (see for example
q1.5\_s0.5x\_0).

These results suggest that convergence of the orbital phase evolution
depends sensitively on the exact parameters of the system under study.
Further investigation of the parameter space is warranted.

\subsection{Impact of PN spin truncation}
\label{sec:ChangeSpinTruncation}
\begin{figure}
  \includegraphics[width=0.96\linewidth]{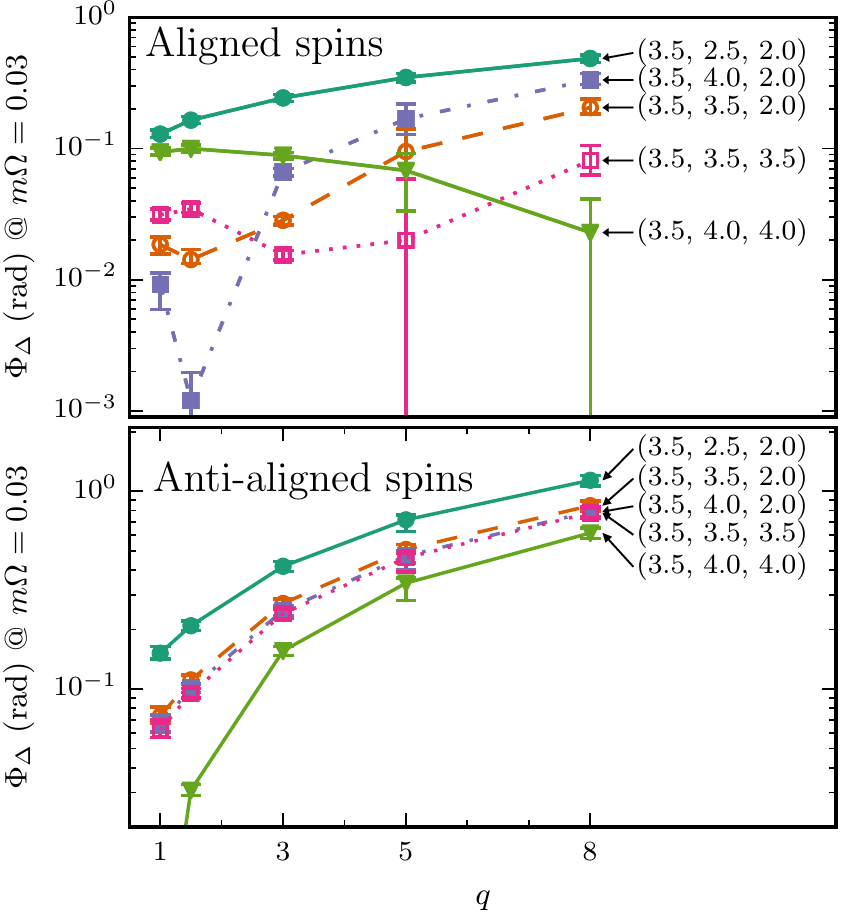}
  \caption{Impact of different choices for spin truncation on orbital
    phase difference $\Phi_{\Delta}$, as a function of mass ratio.
    The lines are labeled by the truncation types, as explained in the
    text.  The upper panel shows all cases for which the spins are
    aligned with the orbital angular momentum; the lower panel shows
    the anti-aligned cases.}
  \label{fig:spinTruncationSummary}
\end{figure}

As mentioned in Sec.~\ref{sec:PNOrders}, post-Newtonian expansions are
not fully known to the same orders for spin and non-spin terms.  Thus,
for example, the expression for flux $\mathcal{F}$ is complete to 3.5
PN order for non-spinning systems, but spinning systems may involve
unknown terms at 2.5 PN order; a similar statement holds for $dE/dx$.
This means that when the ratio in Eq.~\eqref{eq:OmEv}, $\mathcal{F} /
(dE/dx)$, is re-expanded as in the T4 approximant, known terms will
mix with unknown terms.  It is not clear, \textit{a priori}, how such
terms should be handled when truncating that re-expanded series.

Here we examine the effects of different truncation strategies.  We
focus on the Taylor T4 approximant while considering various possible
truncations of the re-expanded form of $\mathcal{F} / (dE/dx)$.  We
denote these possibilities by the orders of (1) the truncation of
non-spin terms, (2) the truncation of spin-linear terms, and (3) the
truncation of spin-quadratic terms.  Thus, for example, in the case
where we keep non-spin terms to 3.5 PN order, keep spin-linear terms
to 2.5 PN order, and keep spin-quadratic terms only to 2.0 PN order,
we write (3.5, 2.5, 2.0).  We consider the following five
possibilities:\\
\hspace*{1em}(i)\,\, (3.5, 3.5, 3.5)\\
\hspace*{1em}(ii)\, (3.5, 4.0, 4.0)\\
\hspace*{1em}(iii) (3.5, 2.5, 2.0)\\
\hspace*{1em}(iv) (3.5, 3.5, 2.0)\\
\hspace*{1em}(v)\, (3.5, 4.0, 2.0).
 
  To increase the impact of the spin-orbit terms, we examine aligned
  and anti-aligned cases from Table~\ref{tbl:Parameters}, with results
  presented in Fig.~\ref{fig:spinTruncationSummary}. For aligned
  cases, no one choice of spin truncation results in small differences
  across all mass ratios.  All choices of spin truncation excepting
  (3.5, 4.0, 4.0) have increasing errors with increasing mass ratio.
  Truncating spin corrections at 2.5 PN order (3.5, 2.5, 2)
  consistently results in the worst matches.  On the other hand, we
  find that, for anti-aligned runs, adding higher order terms always
  improves the match, keeping all terms yields the best result, and
  all choices of truncation give errors which are monotonically
  increasing with mass ratio. Overall, anti-aligned cases have larger
  values of $\Phi_{\Delta}$ when compared to cases with same mass
  ratios. This result is consistent with findings by Nitz
  et~al.~\cite{Nitz:2013mxa} for comparisons between TaylorT4 and
  EOBNRv1 approximants.

\subsection{Further numerical considerations}
\label{sec:tech_cons}

\subsubsection{Numerical truncation error}

\begin{figure}
  \includegraphics[width=0.96\linewidth,trim=0 26 0 0]{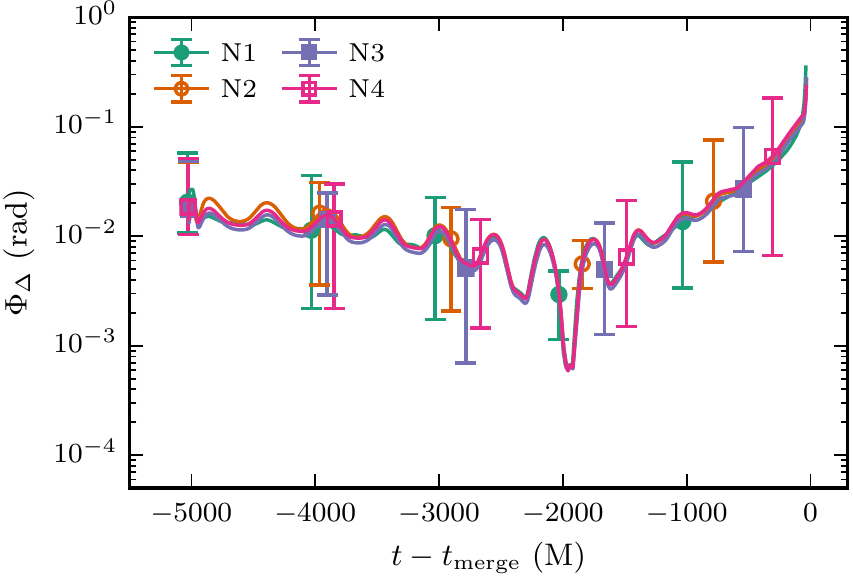}
  \includegraphics[width=0.96\linewidth]{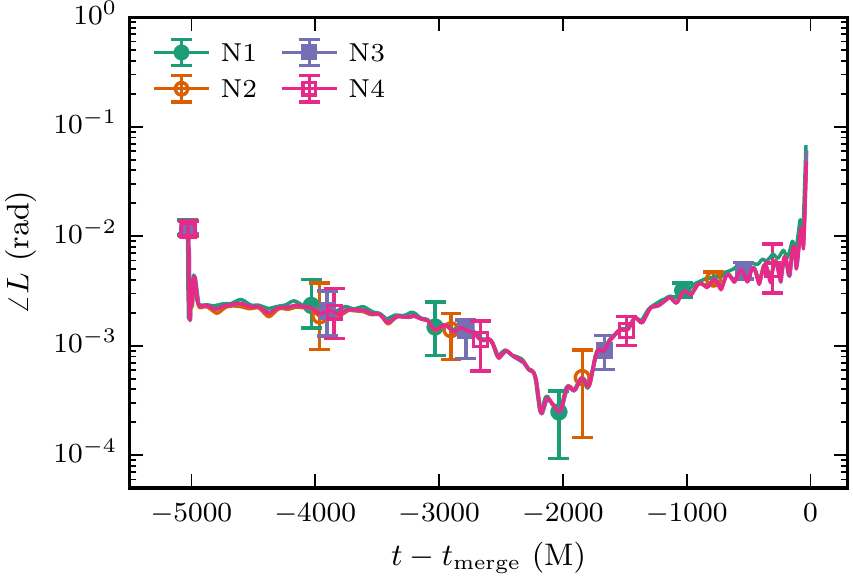}
  \caption{Convergence test with the numerical resolution of the NR
    simulation q1\_twospins. {\bf Top panel}: $\Phi_\Delta$
      with comparisons done at different resolutions. All the curves
      lie within uncertainties due to the matching procedure,
      indicating that numerical truncation error does is not important
      in this comparison. The difference between each curve and the
      highest resolution are of order 15\% and are within the matching
      uncertainties. {\bf Bottom panel}: $\angle L$ with comparisons
      done at all the resolutions. The curves lie within the matching
      uncertainties.} 
  \label{fig:convergenceWithLev}
\end{figure}

Still to be addressed is the effect of the resolution of NR
simulations in the present work.  The simulation q1\_twospins is
available at four different resolutions labeled N1, N2, N3 and N4.  We
match each of these four numerical resolutions with the Taylor T4
approximant, and plot the resulting phase differences $\Phi_\Delta$ in
Fig.~\ref{fig:convergenceWithLev} as the data with symbols and
error bars (recall that the error bars are obtained from the 12
different matching regions we use, cf. Fig.~\ref{fig:avProcedure}).
All four numerical resolutions yield essentially the same
$\Phi_\Delta$.  We furthermore match the three lowest numerical
resolutions against the highest numerical resolution N4 and compute
the phase difference $\Phi_\Delta$.  The top panel of Figure
\ref{fig:convergenceWithLev} shows $\Phi_{\Delta}$ computed with
these 4 different numerical resolutions. All the curves lie on top of
each other and the differences between them are well within the
uncertainties due to the matching procedure. The bottom panel shows
the differences in $\Phi_{\Delta}$ between the highest resolution and
all others. Throughout most of the inspiral, the difference is $\sim
10 \%$.  Similar behavior is observed in other cases where multiple
resolutions of NR simulations are available. We therefore conclude
that the effects of varying numerical resolution do not impact our
analysis.

\subsubsection{Numerical gauge change}
\label{sec:Gauge}

\begin{figure}
  \centering
    \includegraphics[width=0.98\linewidth]{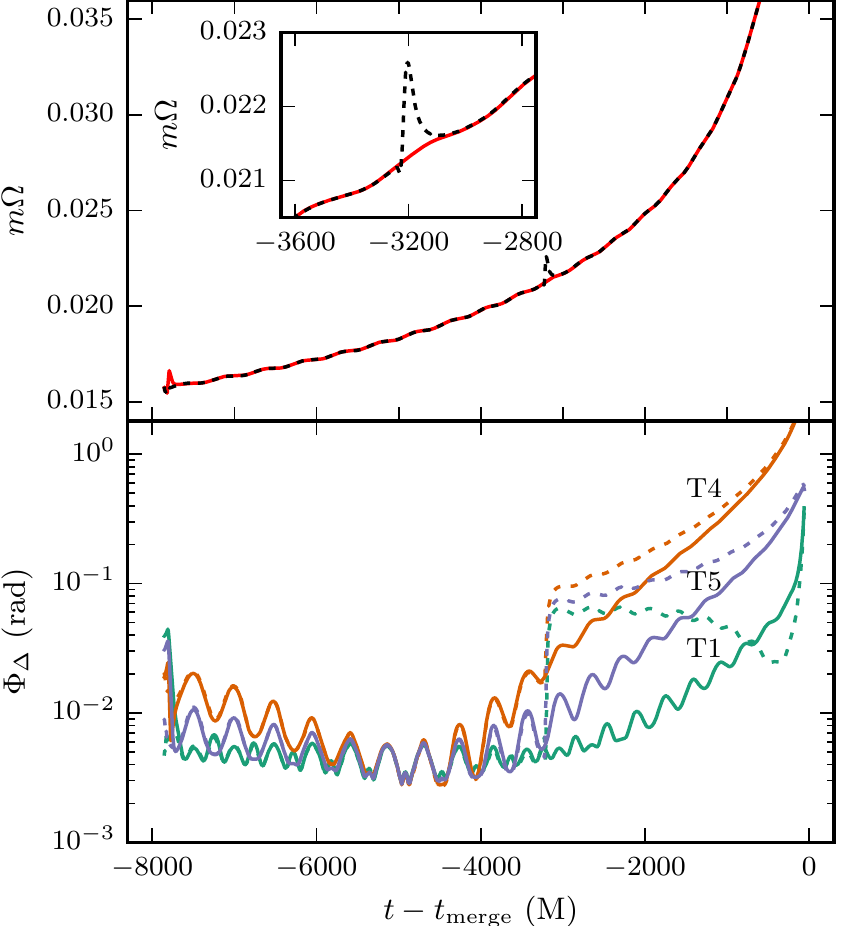}
    \caption{Gauge change during numerical simulation q5\_s0.5x.  The
      solid curves represent the recent re-run of q5\_0.5x that is
      analyzed in the rest of this paper.  The dashed curves represent
      an earlier run SXS:BBH:0058 which changes the gauge at $t-t_{\rm
        merge}\approx -3200M$.  {\bf Top}: behavior of the orbital
      frequency $m\Omega$ in evolution with (dashed curve) and without
      gauge change (solid curve). {\bf Bottom:} $\Phi_{\Delta}$ for
      all Taylor approximants. To avoid matching during the gauge
      change, the matching was done with $m\Omega_{c}=0.017$. }
  \label{fig:gaugeProblems}
\end{figure}

The simulation SXS:BBH:0058 in the SXS catalog uses identical BBH
parameters than q5\_0.5x, but suffers from two deficiencies,
exploration of which will provide some additional insights.  First,
the switch from generalized harmonic gauge with \emph{fixed}
gauge-source functions~\cite{Boyle2007} to \emph{dynamical}
gauge-source functions~\cite{Lindblom2009c,Szilagyi:2009qz} happens
near the middle of the inspiral, rather than close to merger as for
the other simulations considered. This will give us an opportunity to
investigate the impact of such a gauge change, the topic of this
subsection.  Second, this simulation also used too low resolution in
the computation of the black hole spin during the inspiral, which we
will discuss in the next subsection.  We emphasize that the
comparisons presented above did not utilize SXS:BBH:0058, but rather a
re-run with improved technology.  We use SXS:BBH:0058 in this section
to explore the effects of its deficiencies.

While the difference between PN and NR gauges does not strongly impact
the nature of the matching results, a gauge change performed during
some of the runs \emph{does} result in unphysical behavior of physical
quantities such as the orbital frequency. Figure
\ref{fig:gaugeProblems} demonstrates this for case q5\_s0.5x.  The old
run SXS:BBH:0058 with the gauge change exhibits a bump in the orbital
frequency (top panel), which is not present in the re-run (solid
curve).  When matching both the old and the new run to PN, and
computing the phase difference $\Phi_\Delta$, the old run exhibits a
nearly discontinuous change in $\Phi_{\Delta}$ (bottom panel, dashed
curves) while no such discontinuity is apparent in the re-run.

\begin{figure}
\hfill    \includegraphics[scale=0.985775, trim=0 26 0 0]{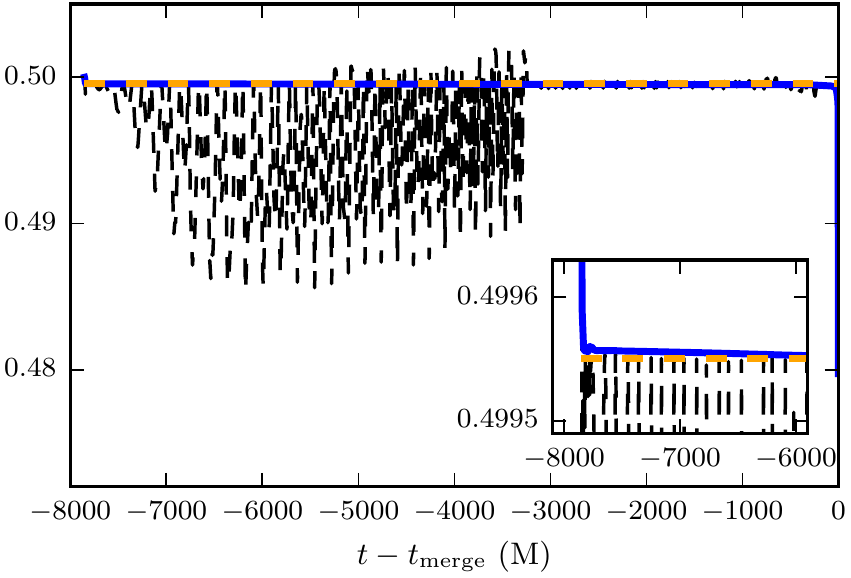}\\
\hfill    \includegraphics[trim=0 26 0 0]{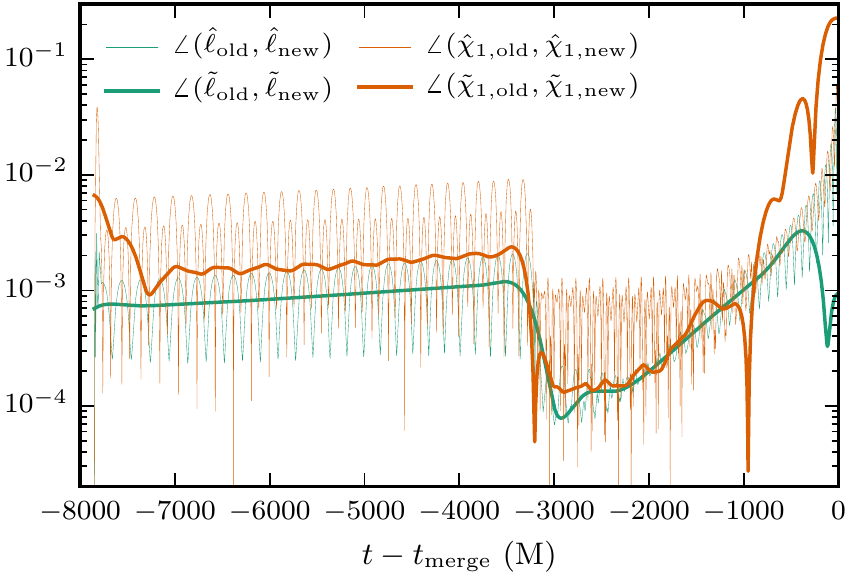}\\
\hfill\includegraphics[scale=1]{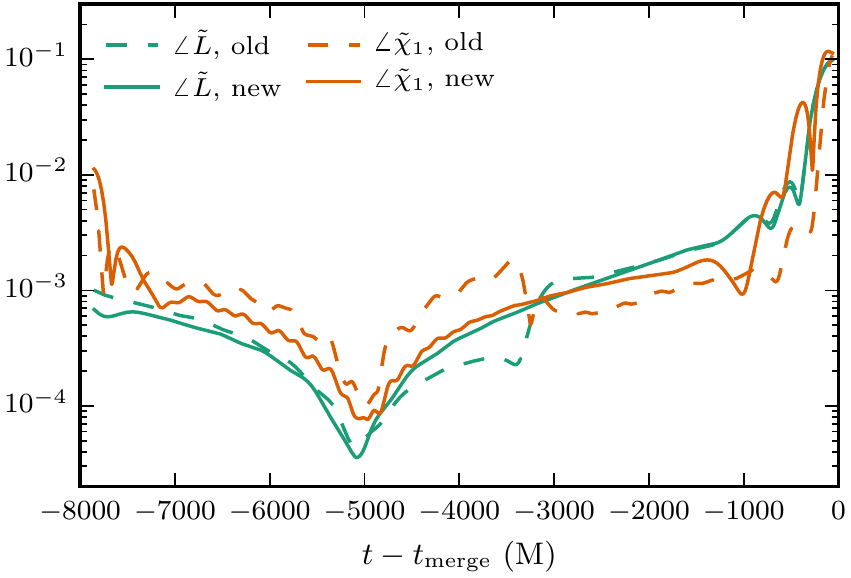}
\caption{{\bf Top}: The magnitude of the spin as a function of time in
  the original run (black) and the new run (blue) as well as the value
  computed with the procedure described in the text (orange). {\bf
    Middle panel}: angles between the spins and normals to the orbital
  plane (thin curves) and their averaged values (bold curves) for the
  original run and the re-run.  {\bf Lower panel}: $\angle
  \tilde{\chi}_{1}$ and $\angle \tilde{\ell}$ for both the old run and
  the re-run (the data of this panel are averaged over 12 matching
  intervals, cf. Fig.~\ref{fig:avProcedure}). To avoid matching during
  the gauge change, the matching was done with $m\Omega_{c}=0.017$.}
    \label{fig:q5SpinComparison}
  \end{figure}

\subsubsection{Problems in quasi-local quantities}
\label{sec:AH-resolution}

Computation of the quasi-local spin involves the solution of an
eigenvalue problem on the apparent horizon followed by an integration
over the apparent horizon,
cf.~\cite{Lovelace2008,OwenThesis,Cook2007}.  In the simulations
q1.0\_0.5x, q1.5\_0.5x and q3.0\_0.5x and in SXS:BBH:0058
(corresponding to q5\_0.5x), too low numerical resolution was used for
these two steps.  While the evolution itself is acceptable, the
extracted spin shows unphysical features.  Most importantly, the
reported spin magnitude is not constant, but varies by several per
cent.  Figure~\ref{fig:q5SpinComparison} shows as example $\chi_1$
from SXS:BBH:0058.  For $t-t_{\rm merge}\le 3200M$ oscillations are
clearly visible.  These oscillations vanish at $t-t_{\rm merge}\approx
3200M$, coincident with a switch to damped harmonic gauge
(cf. Sec.~\ref{sec:Gauge}).  Similar oscillations in q3\_0.5 disappear
when the resolution of the spin computation is manually increased
about 1/3 through the inspiral, without changing the evolution gauge.
Our new re-run q5\_0.5x (using damped harmonic gauge throughout), also
reports a clean $\chi_1$, cf. Fig.~\ref{fig:q5SpinComparison}.  Thus,
we conclude that the unphysical variations in the spin magnitude are
only present if \emph{both} the resolution of the spin computation is
low, and the old gauge conditions of constant $H_a$ are employed.

The NR spin magnitude is used to initialize the PN spin magnitude,
cf. Eq.~\eqref{eq:conserved}.  Therefore, an error in the calculation
of the NR spin would compromise our comparison with PN.  For the
affected runs, we correct the spin reported by the quasi-local spin
computation by first finding all maxima of the spin-magnitude $\chi$
between $500M$ and $2000M$ after the start of the numerical
simulation.  We then take the average value of $\chi$ at those maxima
as the corrected spin-magnitude of the NR simulation.
Figure~\ref{fig:q5SpinComparison} shows the case q5\_0.5x as well as
the rerun described in Sec. \ref{sec:Gauge}. It is evident that this
procedure produces a spin value which is very close to the spin in the
rerun where the problematic behavior is no longer present. Thus, we
adopt it for the three cases where an oscillation in the spin
magnitude is present.

The nutation features shown in Fig.~\ref{fig:ProjectionSpinsQ5} are
qualitatively similar for all our simulations, independent of
resolution of the spin computation and evolution gauge.  When the spin
is inaccurately measured, the nutation trajectory picks up extra
modulations, which are small on the scale of
Fig.~\ref{fig:ProjectionSpinsQ5} and do not alter the qualitative
behavior.

The lower two panels of Fig.~\ref{fig:q5SpinComparison} quantify the
impact of inaccurate spin measurement on the precession-dynamics
comparisons performed in this paper: The middle panel shows the
differences between the spin directions in the original 0058 run and
our re-run q5\_0.5x.  The spin directions differ by as much as 0.01
radians.  However, as the lower panel shows, this difference can
mostly be absorbed by the PN matching, so that $\angle\chi_1$ and
$\angle L$ are of similar magnitude of about $10^{-3}$ radians.

\section{Discussion}
\label{sec:discussion}

We have presented an algorithm for matching PN precession dynamics to
NR simulations which uses constrained minimization.  Using this
algorithm, we perform a systematic comparison between PN and NR for
precessing binary black hole systems. The focus of the comparison is
black hole dynamics only, and we defer discussion of waveforms to
future work. By employing our matching procedure, we find excellent
agreement between PN and NR for the precession and nutation of the
orbital plane. The normals to the orbital plane generally lie within
$10^{-2}$ radians, cf. Fig.~\ref{fig:AngleLAll}. Moreover, nutation
features on the orbital time-scale also agree well between NR and PN,
cf. Fig.~\ref{fig:ProjectionEllHatQ5}.

For the black hole spin direction, the results are less uniform.  The
NR spin direction $\hat\chi_1^{\rm NR}$ shows nutation features that
are qualitatively different than the PN nutation features,
cf. Fig.~\ref{fig:ProjectionSpinsQ5}.  The disagreement in nutation
dominates the agreement of $\hat \chi_1^{\rm NR}$ with
$\hat\chi_1^{\rm PN}$; averaging away the nutation features
substantially improves agreement,
cf. Fig.~\ref{fig:AngleSmoothedS2cases}.  The orbit-averaged
spin directions agree with PN to the same extent that the $\hat\ell$
direction does (with and without orbit averaging),
cf.~Fig.~\ref{fig:AngleLAll}.

Turning to the convergence properties of PN, we have performed PN-NR
comparisons at different PN order of the precession equations.  For
both orbital angular momentum $\hat\ell$ and the spin direction
$\hat\chi_1$, we observe that the convergence of the PN results toward
NR is fast and nearly universally monotonic, cf.
Fig.~\ref{fig:PrecPNOrder}.  At the highest PN orders, the spin
results might be dominated by the difference in nutation features
between PN and NR.

The good agreement between PN and NR precession dynamics are promising
news for gravitational wave modeling.  Precessing waveform models
often rely on the post-Newtonian precession equations,
e.g.~\cite{Arun:2008kb,Hannam:2013oca}.  Our results indicate that the
PN precession equations are well suited to model the precessing frame,
thus reducing the problem of modeling precessing waveforms to the
modeling of orbital phasing only.

The accuracy of the PN orbital phase evolution, unfortunately, does
not improve for precessing systems. Rather, orbital phasing errors are
comparable between non-precessing and precessing configurations, \
cf. Fig.~\ref{fig:q3_OmegaConv}.  Moreover, depending on mass-ratio
and spins, some Taylor approximants match the NR data particularly
well, whereas others give substantially larger phase differences,
cf. Fig.~\ref{fig:massRatioDep}.  This confirms previous
work~\cite{Damour:2010,Hannam:2010,Hannam:2010,Santamaria:2010yb,MacDonald:2012mp,MacDonald:2011ne}
that the PN truncation error of the phase evolution is important for
waveform modeling.

We have also examined the effects of including
partially known spin contributions to the evolution of the orbital
frequency for the Taylor T4 approximant. For aligned runs, including
such incomplete information usually improves the match, but the
results are still sensitive to the mass ratio of the binary (top panel
of Fig~\ref{fig:spinTruncationSummary}). For anti-aligned runs, it
appears that incomplete information always improves the agreement of
the phasing between PN and NR (bottom panel of
Fig~\ref{fig:spinTruncationSummary}).

In this work we compare gauge-dependent quantities, and thus must
examine the impact of gauge choices on the conclusions listed above.
We consider it likely that the different nutation features of
$\hat\chi_1$ are determined by different gauge choices.  We have also
seen that different NR gauges lead to measurably different evolutions
of $\hat\chi$, $\hat\ell$, and the phasing,
cf. Fig.~\ref{fig:gaugeProblems} and~\ref{fig:q5SpinComparison}.  We
expect, however, that our conclusions are fairly robust to the
gauge ambiguities for two reasons. First, in the matched PN-NR
comparison, the impact of gauge differences is quite small, cf. lowest
panel of Fig.~\ref{fig:q5SpinComparison}.  Second, the near universal,
monotonic, and quick convergence of the precession dynamics with
precession PN order visible in Fig.~\ref{fig:PrecPNOrder} would not be
realized if the comparison were dominated by gauge effects.  Instead,
we would expect PN to converge to a solution {\em different} from the
NR data.

\begin{acknowledgments}
  We thank Kipp Cannon, Francois Foucart, Prayush Kumar, Abdul Mrou\'e
  and Aaron Zimmerman for useful discussions.  Calculations were
  performed with the {\tt SpEC}-code~\cite{SpECwebsite}.  We
  gratefully acknowledge support from NSERC of Canada, from the Canada
  Research Chairs Program, and from the Canadian Institute for
  Advanced Research.  We further gratefully acknowledge support from
  the Sherman Fairchild Foundation; from NSF Grants PHY-1306125 and
  AST-1333129 at Cornell; and from NSF Grants No. PHY-1440083 and
  AST-1333520 at Caltech.  Calculations were performed at the GPC
  supercomputer at the SciNet HPC Consortium~\cite{scinet}; SciNet is
  funded by: the Canada Foundation for Innovation (CFI) under the
  auspices of Compute Canada; the Government of Ontario; Ontario
  Research Fund (ORF) -- Research Excellence; and the University of
  Toronto.  Further computations were performed on the Zwicky cluster
  at Caltech, which is supported by the Sherman Fairchild Foundation
  and by NSF award PHY-0960291; and on the NSF XSEDE network under
  grant TG-PHY990007N.
\end{acknowledgments}

\appendix 

\section{Post-Newtonian dynamics}
\label{ap:PN}

We consider compact object binary with masses $m_{1,2}$ and carrying
angular momentum $\vec{S}_{1,2}$. The post-Newtonian expressions are
most conveniently written using the following symbols:

\begin{align}
m &= m_{1}+m_{2},\\
\nu &= \frac{m_{1}m_{2}}{m^{2}},\\
\delta &= \frac{m_{1}-m_{2}}{m},\\
\vec{S} &= \vec{S}_{1}+\vec{S}_{2},\\
s_{l} &= \frac{\vec{S}\cdot\hat\ell}{m^{2}},\\
s_{n} &= \frac{\vec{S}\cdot\hat n}{m^{2}},\\
\vec{\Sigma} &= \frac{m}{m_2}\vec{S}_{2} - \frac{m}{m_1}\vec{S}_{1},\\
\sigma_{l} &=\frac{\vec\Sigma\cdot\hat\ell}{m^{2}}, \\
\sigma_{n} &= \frac{\vec\Sigma\cdot\hat n}{m^{2}}, \\
\vec{\chi}_{s} &= \frac{1}{2}\left(\vec{\chi}_{1}+\vec{\chi}_{2}\right),\\
\vec{\chi}_{a} &= \frac{1}{2}\left(\vec{\chi}_{1}-\vec{\chi}_{2}\right),\\
\vec{S}_0 & = \frac{m}{m_1}\vec{S}_{1}+\frac{m}{m_2}\vec{S}_{2},\\
\vec{s}_0 & = \frac{\vec{S}_0}{m^2}.
\end{align}

\subsection{Energy and Flux}
\label{ap:energyFlux}

The energy and flux are written as power series in the expansion
parameter $x\equiv (m\Omega)^{2/3}$:

\begin{eqnarray}
E(x) &= -\frac{1}{2}\,m\nu x\,\left(1+\sum_{k=2}a_{k}x^{k/2}\right), \label{eq:EExpansion} \\
\mathcal{F}(x) &= \frac{32}{5}\nu^{2}x^{5}\left(1+\sum_{k=2}
  b_{k}x^{k/2}\right).
\label{eq:FExpansion}
\end{eqnarray}
For the energy, coefficients are given explicitly by:
\begin{widetext}
\begin{eqnarray}
a_{2} &=& -\frac{3}{4} - \frac{\nu}{12}, \\
a_{3} &=&  2\delta\sigma_{l}+\frac{14}{3}s_{l}, \\
a_{4} &=& -\frac{27}{8} + \frac{19}{8}\nu - \frac{1}{24}\nu^{2} +
\nu{(\vec{\chi}_{s}^{2}-\vec{\chi}_{a}^{2} -
  3[(\vec{\chi}_{s}\cdot\ellHat)^{2} -
  (\vec{\chi}_{a}\cdot\ellHat)^{2}]} \nonumber \\
  &&+ (\frac{1}{2}-\nu)\{ \vec{\chi}_{s}^{2}+\vec{\chi}_{a}^2-3[(\vec{\chi}_{s}\cdot\ellHat)^{2} + (\vec{\chi}_{a}\cdot\ellHat)^{2}]\}+\delta\{\vec{\chi}_{s}\cdot\vec{\chi}_{a} - 3[(\vec{\chi}_{s}\cdot\ellHat)(\vec{\chi}_{a}\cdot\ellHat)]\},
\label{eq:Energy-a4}\\
a_{5} &=& 11s_{l} + 3\delta\sigma_{l} + \nu\left[-\frac{61}{9}s_{l} - \frac{10}{3}\delta\sigma_{l}\right],\\
a_{6} & =& -\frac{675}{64} + \left[\frac{34445}{576}-\frac{205}{96}\pi^{2}\right]\nu - \frac{155}{96}\nu^{2}-\frac{35}{5184}\nu^{3},\\
a_{7} &=& \left(\frac{135}{4} - \frac{367}{4}\nu + \frac{29}{12}\nu^{2}\right)s_{l} + \delta\left(\frac{27}{4}-39\nu+\frac{5}{4}\nu^{2}\right)\sigma_l.
\end{eqnarray}
Meanwhile for the flux $\mathcal{F}$:
\begin{eqnarray}
b_{2} &=& -\frac{1247}{336}-\frac{35}{12}\nu, \\
b_{3} &=& 4\pi - 4s_l - \frac{5}{4}\delta\sigma_{l}, \\
b_{4} &=& -\frac{44711}{9072} + \frac{9271}{504}\nu + \frac{65}{18}\nu^{2}  + \left(\frac{287}{96}+\frac{\nu}{24}\right)(\vec{\chi}_{s}\cdot\ellHat)^{2}\nonumber\\
&&- \left(\frac{89}{96}+\frac{7\nu}{24}\right)\vec{\chi}_{s}^{2} + \left(\frac{287}{96}-12\nu\right)(\vec{\chi}_{a}\cdot\ellHat)^{2} + \left(-\frac{89}{96}+4\nu\right)\vec{\chi}_{a}^{2} + \frac{287}{48}\delta(\vec{\chi}_{s}\cdot\ellHat)(\vec{\chi}_{a}\cdot\ellHat) - \frac{89}{48}\delta(\vec{\chi}_{s}\cdot\vec{\chi}_{a}),
\label{eq:Flux-b4}
\\
b_{5} &=& -\frac{8191}{672}\pi - \frac{9}{2}s_{l} - \frac{13}{16}\delta\sigma_{l} + \nu\left[-\frac{583}{24}\pi + \frac{272}{9}s_{l} + \frac{43}{4}\delta\sigma_{l}\right],\\
b_{6} &=& \frac{6643739519}{69854400} + \frac{16}{3}\pi^{2} - \frac{1712}{105}\gamma_{E} - \frac{856}{105}\log(16x)+
 \left(\frac{-134543}{7776}+\frac{41}{48}\pi^{2}\right)\nu -
 \frac{94403}{3024}\nu^{2}-\frac{775}{324}\nu^{3} \nonumber\\
 && - 16\pi s_{l} - \frac{31\pi}{6}\delta\sigma_{l},\\
 b_{7}&=& \left(\frac{476645}{6804} + \frac{6172}{189}\nu -
   \frac{2810}{27}\nu^{2}\right)s_{l} +
 \left(\frac{9535}{336}+\frac{1849}{126}\nu -
   \frac{1501}{36}\nu^{2}\right)\delta\sigma_{l} \nonumber\\
&&+  \left(-\frac{16285}{504} \frac{214745}{1728}\nu +\frac{193385}{3024}\nu^{2}\right)\pi,\\
 b_{8} &=& \left(-\frac{3485\pi}{96} + \frac{13879\pi}{72}\nu\right)s_{l} + \left(-\frac{7163\pi}{672} + \frac{130583\pi}{2016}\nu\right)\delta\sigma_l,
\end{eqnarray}
where $\gamma_E$ denotes Euler's constant.

\subsection{Precession dynamics}
\label{ap:orbitalEvolution}
The evolution of the orbital plane is governed by the frequency
$\varpi$ in Eq.~\eqref{eq:ellHatEv}, which is defined in terms of two
auxiliary quantities, $\gamma=m/r$ and
$a_l=\vec{a}\cdot{\ellHat}$:

\begin{eqnarray}
  \gamma & = & x \bigg\{1 + \frac{3-\nu}{3}\,x
    +\frac{3\sigma_l+5s_l}{3}\,x^{3/2}
    + \frac{12-65\nu}{12}\,x^{2}
    + \left(\frac{30+8\nu}{9}s_l+2\sigma_l\delta\right)\;x^{5/2} \nonumber \\
  &&\quad 
    +\left[1 + \nu \left(- \frac{2203}{2520}-\frac{41 \pi^{2}}{192}\right) 
      + \frac{229 \nu^{2}}{36} + \frac{\nu^{3}}{81}\right]\,x^3
+ \left( \frac{60-127\nu-72\nu^2}{12}
\,s_{l} + 
\frac{16-61\nu-16}{6}
\sigma_{l} \delta 
\right)\,x^{7/2}
\nonumber \\    
&& \quad +x^{2} \left(\vec{s}_{0}^{\,2} - 3 (\vec s_0\cdot\vec\ell)^{2}\right)\bigg\},
  \label{eq:gammaPN}\\
%
  a_l&=&  \frac{x^{\frac{7}{2}}}{m}
  \left\{7 s_{n} + 3 \sigma_{n} \delta +  x \left[s_{n} \left(- \frac{29
          \nu}{3} - 10\right) + \sigma_{n} \delta \left(- \frac{9
          \nu}{2} - 6\right)\right] \right. \nonumber \\
  && + \left. x^{2} \left[s_{n} \left(\frac{52
          \nu^{2}}{9} + \frac{59 \nu}{4} + \frac{3}{2}\right) +
      \sigma_{n} \delta \left(\frac{17 \nu^{2}}{6} + \frac{73 \nu}{8} +
        \frac{3}{2}\right)\right] \right\}- \frac{3x^{4}}{m} (\vec{s}_{0}\cdot\ellHat)(\vec{s}_{0}\cdot\nHat).
  \label{eq:al}
\end{eqnarray}
Note that we have dropped the pure gauge term
$-\frac{22}{3}\ln\left(r/r_{0}'\right)$ from $\gamma$.  We now have
\begin{equation}
\label{eq:varpi}
\varpi = \frac{a_l\,\gamma}{x^{3/2}}.
\end{equation}
The spins obey Eqs.~\eqref{eq:SpinsEv} with

\begin{eqnarray}
\label{eq:OmegaSpinsEv}
\vec{\Omega}_{1} &=&\ellHat\frac{x^{\frac{5}{2}}}{m} \Bigg\{
\frac{-3\delta+2\nu+3}{4}
+ x \left[\frac{10\nu-9}{16}\delta
 - \frac{\nu^2}{24}
    + \frac{5\nu}{4} + \frac{9}{16}\right]
  + x^{2} \left[
\frac{-5\nu^2+156\nu-27}{32}\delta
- \frac{\nu^3}{48} - \frac{105\nu^2}{32} + \frac{3\nu}{16}+
    \frac{27}{32}\right] \Bigg\} \nonumber \\
&&+ \frac{x^{3}}{m^3} \left[\frac{3m_{1}^{2}}{q}(\vec{\chi}_{1}\cdot\nHat) \nHat - m_2^{2}\vec{\chi}_{2} + 3m_{2}^{2}(\vec{\chi}_{2}\cdot\nHat)  \nHat\right].
\end{eqnarray}

The expression for $\vec{\Omega}_{2}$ is obtained by
$\vec{\chi}_{1}\leftrightarrow\vec{\chi}_{2}$, $m_{1}\leftrightarrow
m_{2}$, $\delta\leftrightarrow-\delta$ and $q \leftrightarrow 1/q$.

\end{widetext}
We re-expand the right-hand-side of Eq.~\eqref{eq:varpi}, and truncate
the expansion for $\varpi$ and $\vec\Omega_{1,2}$ at the same power of
$x$ \emph{beyond the leading order}.  We refer to the order of the
last retained terms as the \emph{precession PN order}.  For the
majority of comparisons presented in this paper, we truncate at 3.5PN;
truncation at lower PN order is only used in
Sec.~\ref{sec:PNConvergence}. Note that spin-squared interactions
imply the lack of circular orbits for generic orientations of the
spins. We neglect these complications in the present work.

\section{Useful quaternion formulas}
\label{sec:UsefulQuaternionFormulas}

We refer the reader to other sources~\cite{DoranLasenby:2003,
  Boyle:2013} for general introductions to quaternions.  Here, we
simply give a few formulas that are particularly important in this
paper.  First, we introduce some basic notation to be used for the
four components of a general quaternion $Q$:
\begin{equation}
  \label{eq:QuaternionComponents}
  Q = (q_{0}, q_{1}, q_{2}, q_{3}) = q_{0} + \vec{q}~.
\end{equation}
In this notation, the quaternion conjugate is just $\bar{Q} = q_{0} -
\vec{q}$, and we note that the product of quaternions is given by
\begin{equation}
  \label{eq:QuaternionProduct}
  P\, Q = p_{0}\, q_{0} - \vec{p} \cdot \vec{q} + p_{0}\vec{q}+q_{0}\vec{p}+ \vec{p} \times
  \vec{q}~.
\end{equation}
The norm of a quaternion $Q$ is defined by
  $\abs{Q}^2 = Q\, \bar{Q}$.  The inverse of a quaternion is
$Q^{-1} = \bar{Q} / \abs{Q}^{2}$, which means that the inverse of a
unit quaternion is simply its conjugate.  The components of a unit
quaternion $R=r_0+\vec r$ satisfy
$R\, \bar{R} = r_0^2+\vec r\cdot\vec r=1$.  Unit quaternions are
  usually referred to as ``rotors''.  Any rotation can be expressed as
  a rotor, where the rotor acts on a vector $\vec{v}$ according to the
  transformation law
\begin{equation}
  \label{eq:VectorTransformationLaw}
  \vec{v}\,' = R\, \vec{v}\, \bar{R}~.
\end{equation}
The form of this expression ensures that $\vec{v}\,'$ is a pure
  vector; it has zero scalar part.  To see this, we note that a
  quaternion has zero scalar part if and only if its conjugate equals
  its negative, which is true of the right-hand side above.  We can
  use this fact, along with
  $\vec{p} \cdot \vec{q} = -\frac{1}{2} (\vec{p}\, \vec{q} + \vec{q}\,
  \vec{p})$
  and the unit-norm property $R\, \bar{R} = 1$, to see that the
  right-hand side above is indeed an isometry.  Finally, simple
  arguments using the cross product can show that such a
  transformation preserves orientation, and since the origin is fixed,
  it is therefore a simple rotation for any rotor $R$.

\subsection{Exponential, logarithms, and square roots}
\label{sec:Exponentiallogarithmsandsquareroots}
The quaternions are closely analogous to complex numbers, except that
quaternions do not commute in general.  One striking example of this
analogy is Euler's formula, which generalizes quite directly.  If we
define the exponential of a quaternion by the usual power series, we
get for a unit vector $\hat u$
\begin{equation}
  \label{eq:QuaternionExponential}
  \exp[\theta\, \hat{u}] = \cos\theta + \hat{u} \sin\theta~,
\end{equation}
which is precisely Euler's formula with $i$ replaced by $\hat{u}$.
Every rotor $R=r_0+\vec r$ can be expressed in this form, so it is
easy to see that the logarithm of any rotor has zero scalar part and
is given by
\begin{equation}
  \label{eq:RotorLogarithm}
  \vec{\mathfrak{r}} \defined \log R = \frac{\vec{r}}
  {\abs{\vec{r}\,}}\, \arctan \frac{\abs{\vec{r}\,}} {r_{0}}~.
\end{equation}
It is useful to note that the logarithm of a rotor is parallel to the
vector part of the rotor.  Finding the magnitude of $\vec{r}$, of
course, is just the usual square root of the sum of the squares of its
components.  And the $\arctan$ function is applied to real values, so
we can use standard implementations of the \texttt{atan2} function to
evaluate it.  So we see that both the exponential and logarithm of
quaternions are extremely simple and numerically robust to calculate.

These formulas can also be used to define general powers of
quaternions.  For the purposes of this paper, however, we only need
one particular power of a quaternion: the square root.  More
specifically, given two unit vectors $\hat{u}$ and $\hat{w}$, we need
the rotor that takes $\hat{w}$ to $\hat{u}$ by the smallest rotation
possible, which is a rotation in their common plane.  This rotor is
given~\cite{Boyle:2013} by
\begin{equation}
  \label{eq:RotorSquareRoot}
  R_{\hat{w} \to \hat{u}} = \sqrt{-\hat{u}\, \hat{w}} = \pm \frac{1 -
    \hat{u}\, \hat{w}} {\sqrt{2[1-(\hat{u}\, \hat{w})_{0}]}}~.
\end{equation}
In this expression, $\hat u\,\hat w$ represents the result of
quaternion multiplication of the quaternions $\hat u$ and $\hat w$.
$(\hat u\,\hat w)_0$ represents the scalar part of this product, so
that the square root in the denominator is acting on a real number.
The sign ambiguity is generally irrelevant because of the double-sided
transformation law for vectors,
Eq.~\eqref{eq:VectorTransformationLaw}.  However, in certain special
applications such as rotor interpolation, the sign must be chosen
carefully to be continuous~\cite{Boyle:2013}.

\subsection{Deriving the frame rotor from
  \texorpdfstring{$\ellHat$}{ellHat} and
  \texorpdfstring{$\nHat$}{nHat}}
\label{sec:Derivingtheframerotorfromellhatandnhat}
For both numerical relativity simulations and Post-Newtonian
evolutions we have data about the positions and velocities of the
black holes, that can be used to derive the frame rotor $\Rf$,
cf. Fig.~\ref{fig:orbitalDefns}.  Given positions of the black holes
as functions of time, it's a simple matter to calculate their unit
separation vector $\nHat$, and then to calculate $\ellHat$ using
\begin{equation}
  \label{eq:ellHatFromnHat}
  \Omega\, \ellHat = \hat{n} \times \dot{\hat{n}}~.
\end{equation}
Going from $\ellHat$ and $\nHat$ to the frame rotor $\Rf$, the idea is
to first rotate $\hat{z}$ onto $\ellHat$.  This will also rotate
$\hat{x}$ onto some $\hat{x}'$.  We then need to rotate $\hat{x}'$
onto $\nHat$, while leaving $\ellHat$ in place.  Of course, the
$\nHat$-$\hat{x}'$ is orthogonal to $\ellHat$, so we just perform a
rotation in that plane.  This is easily accomplished by the following
formula:
\begin{subequations}
  \label{eq:FrameRotorFromBasisVectors}
  \begin{gather}
    R_{i} = \sqrt{-\ellHat\, \hat{z}\,}~, \\[0.5em]
    \Rf = \sqrt{-\nHat\, (R_{i}\, \hat{x}\, \bar{R}_{i})\,}\, R_{i}~.
  \end{gather}
\end{subequations}
Again, the square roots are to be evaluated using
Eq.~\eqref{eq:RotorSquareRoot}.

\subsection{Comparing frame rotors}
\label{sec:ComparingFrames}
Reference~\cite{Boyle:2013} introduced a simple, geometrically
invariant measure $R_{\Delta}$ that encodes the difference between two
precessing systems as a function of time, easily reduced to a single
real number $\Phi_{\Delta}$ expressing the magnitude of that
difference.  These quantities were mentioned in
Sec.~\ref{CharacterizingPrecessionByRotors} without much motivation,
here we briefly review that motivation.

In general, we assume that there are two (analytical or numerical)
descriptions of the same physical system, and that we have two
corresponding frames $\RfA$ and $\RfB$.  To understand the difference
between the frames, we can simply take the rotation that takes one
frame onto the other.  In this case, the rotor taking frame A onto
frame B is
\begin{equation}
  R_{\Delta} \defined \RfB\, \RbarfA.
\end{equation}
Rotors compose by left multiplication, so it is not hard to see that
this does indeed take $\RfA$ onto $\RfB$ because the inverse of $\RfA$
is just its conjugate, so $R_{\Delta}\, \RfA = \RfB$.

A particularly nice feature of $R_{\Delta}$ is that it is completely
independent of the inertial basis frame
$(\hat{x}, \hat{y}, \hat{z})$ with respect to which we define the
moving frames.  That is, if we have another basis frame
$(\hat{x}', \hat{y}', \hat{z}')$, there is some $R_{\delta}$ such
that $\hat{x}' = R_{\delta}\, \hat{x}\, \bar{R}_{\delta}$, etc.  The
frame rotors would transform as $\RfA \mapsto \RfA' = \RfA\,
R_{\delta}$, in which case we obtain
\begin{equation}
  \RfB'\, \RbarfA' = \RfB\, R_{\delta}\, \bar{R}_{\delta}\, \RbarfA =
  \RfB\, \RbarfA.
\end{equation}
That is, $R_{\Delta}$ is invariant.

Now, we seek a relevant measure of the magnitude of the rotation
$R_{\Delta}$.  We know that it may be written as a rotation through an
angle $\phi$ about an axis $\hat{v}$.  Clearly, $\phi$ is the measure
we seek.  The rotor corresponding to such a rotation is given by
$R = \exp[\phi\, \hat{v}/2]$.  Thus, to find the angle, we just use
the logarithm: $\phi = 2 \lvert \log R \rvert$, where the norm is the
usual vector norm.  Again, the formula for the logarithm of a rotor is
a simple combination of standard trigonometric functions applied to
real numbers, as shown above.  Using this interpretation with our
difference rotor, we see that the appropriate definition is
\begin{equation}
  \Phi_{\Delta} \defined 2 \left\lvert \log \left[ \RfB\, \RbarfA
    \right] \right\rvert~.
\end{equation}
There is information contained in the direction of the logarithm.  For
example, the component along $\ellHat$ is related to the difference in
orbital phase for non-precessing systems, while the component
orthogonal to $\ellHat$ is related to the direction and magnitude of
the difference in $\ellHat$ itself.  For the sake of simplicity,
however, we focus on the magnitude of the logarithm, as given above.

\subsection{Inadequacy of \texorpdfstring{$\Phi_A-\Phi_B$}{PhiA-PhiB}
  for comparisons of precessing  systems}
\label{sec:InadequacyOfSeparatePhases}

We claim that it is impossible---when analyzing precessing
systems---to compare two rotations $R_{A}$ and $R_{B}$ in a
non-degenerate and geometrically invariant way by defining some
phases $\Phi_{A}$ and $\Phi_{B}$ for them separately, and then
comparing them as $\Phi_{A}-\Phi_{B}$.  Here, ``non-degenerate''
  means that the phase difference is zero if and only if $R_{A}$ and
  $R_{B}$ represent the same rotation, and ``geometrically invariant''
  means that the result is not affected by an overall rotation of the
  basis used to define $R_{A}$ and $R_{B}$.  In this section, we
prove that statement.

We begin by defining a function $\Phi$ such that $\Phi(R_A) = \Phi_A$
and $\Phi(R_B) = \Phi_B$.  The domain of this function is a rotation
group, which could be the one-dimensional group $\U(1)$ for
non-precessing systems, but must be the full three-dimensional
group\footnote{Even though it is a double cover of the physical
  rotation group $\SO(3)$, we use $\SU(2)$ here for consistency of
  notation, because it is the group of unit quaternions.  The proof
  would actually be slightly simpler for $\SO(3)$; we would have
  $\Phi(R_{A}) = \Phi(R_{B})$, if and only if $R_{A} = R_{B}$, and
  $\ker\Phi' = \{1\}$.}  $\SU(2)$ for general precessing systems.  The
range of $\Phi$ is the usual range of phases, the additive group of
real numbers modulo $2\pi$.  It will be useful to note that this is
isomorphic to $\U(1)$.  Finally, non-degeneracy is the condition
that $\Phi_{A}-\Phi_{B} = 0$ [or equivalently $\Phi(R_{A}) =
\Phi(R_{B})$] if and only if $R_{A} = \pm R_{B}$.

The condition of geometric invariance can be written as a condition on
$\Phi$ itself.  If, for example, we measure everything with respect to
some basis $(\hat{x}, \hat{y}, \hat{z})$, and then measure again with
respect to some other basis $(\hat{x}', \hat{y}', \hat{z}')$, we
should get the same answer.  Now, there is some rotor $R_{\delta}$
that takes the first basis into the second.  If $R_{A}$ is defined
with respect to the first basis, then the equivalent quantity will be
$R_{A}\, R_{\delta}$ with respect to the second.  Geometric invariance
is then the statement
\begin{equation}
  \label{eq:GeometricInvariance}
  \Phi(R_{A}\, R_{\delta}) - \Phi(R_{B}\, R_{\delta}) = \Phi(R_{A}) -
  \Phi(R_{B}),
\end{equation}
for \emph{any} choice of $R_{\delta}$ in $\SU(2)$.  We will show that
there is no such $\Phi$ because the rotation group $\SU(2)$ is not
isomorphic to $\U(1)$.

Since Eq.~\eqref{eq:GeometricInvariance} is true for \emph{any} rotor
$R_{\delta}$, we can choose $R_{\delta} = R_{B}^{-1}$, and find that
\begin{equation}
  \label{eq:GeometricInvariance_B}
  \Phi(R_{A}\, R_{B}^{-1}) - \Phi(1) = \Phi(R_{A}) - \Phi(R_{B}).
\end{equation}
Now, we define another function $\Phi'(R) = \Phi(R) - \Phi(1)$.  The
last equation becomes
\begin{equation}
  \label{eq:GeometricInvariance_Bprime}
  \Phi'(R_{A}\, R_{B}^{-1}) = \Phi'(R_{A}) - \Phi'(R_{B}).
\end{equation}
In exactly the same way, we can see that
\begin{equation}
  \label{eq:GeometricInvariance_Aprime}
  \Phi'(R_{B}\, R_{A}^{-1}) = \Phi'(R_{B}) - \Phi'(R_{A}) = -
  \Phi'(R_{A}\, R_{B}^{-1}).
\end{equation}
This must be true for \emph{all} values of $R_{A}$ and $R_{B}$, so we
have shown that
\begin{equation}
  \label{eq:Inversion}
  \Phi'(R^{-1}) = - \Phi'(R),
\end{equation}
for arbitrary $R$.  Therefore, we can also see from
Eq.~\eqref{eq:GeometricInvariance_Bprime} that
\begin{equation}
  \label{eq:Homomorphism}
  \Phi'(R_{1}\, R_{2}) = \Phi'(R_{1}) + \Phi'(R_{2}),
\end{equation}
for arbitrary $R_{1}$ and $R_{2}$.  This is precisely the statement
that $\Phi'$ is a homomorphism [from $\SU(2)$ to the additive group of
real numbers modulo $2\pi$].

However, now we can impose the condition that $\Phi_{A}-\Phi_{B} = 0$
if and only if $R_{A} = \pm R_{B}$.  Using the properties of
homomorphism, it is clear that this is equivalent to the statement
that the set of all elements that map to $0$ under $\Phi'$ is
$\ker \Phi' = \{-1,1\}$.
Then, the First Group Isomorphism Theorem~\cite{DummitFoote:1999} says
that the image of $\Phi'$ is isomorphic to $\SU(2)$ modulo this
kernel, which of course is just $\SO(3)$.  But the image of $\Phi'$ is
(possibly a subgroup of) the group $\U(1)$, which is obviously not
isomorphic to $\SO(3)$.%
\footnote{This statement will not be controversial, but for form's
  sake we can prove it simply by noting that $\U(1)$ is commutative,
  whereas we can find elements of $\SO(3)$ that do not commute.} %
Therefore, it is impossible to construct a function fulfilling our
requirements for precessing systems.

It is, however, interesting to note that if our rotation group were
not $\SU(2)$, but the one-dimensional rotation group $\U(1)$, there
would be no contradiction.  This is how it \emph{is} possible to
construct a useful measure of the form $\Phi_{A}-\Phi_{B}$ for
\emph{non}-precessing systems, because the rotations can be restricted
to rotations about the orbital axis.  On the other hand, for
precessing systems, the measure $\Phi_{\Delta}$ described in
Secs.~\ref{CharacterizingPrecessionByRotors}
and~\ref{sec:ComparingFrames} is able to satisfy both key features of
a useful measure (non-degeneracy and geometric invariance) because it
simply does not attempt to define a homomorphism from the rotation
group; rather, it defines a (non-homomorphic but invariant and
non-degenerate) function from \emph{two copies of} the rotation group
onto phases, $\SU(2) \times \SU(2) \to \U(1)$.




\bibliography{References}

\end{document}